\documentclass[prodmode]{acmsmall}

\begin{document}

\markboth{Nikolas Landia et al.}{Deeper Into the Folksonomy Graph}

\title{Deeper Into the Folksonomy Graph: FolkRank Adaptations and Extensions for Improved Tag Recommendations}
\author{NIKOLAS LANDIA
\affil{University of Warwick, UK}
STEPHAN DOERFEL
\affil{University of Kassel, Germany}
ROBERT J\"{A}SCHKE
\affil{L3S Research Center, Germany}
SARABJOT SINGH ANAND
\affil{University of Warwick, UK}
ANDREAS HOTHO
\affil{University of W\"{u}rzburg, Germany}
NATHAN GRIFFITHS
\affil{University of Warwick, UK}
}

\begin{abstract}
The information contained in social tagging systems is often modelled as a graph of connections between users, items and tags. Recommendation algorithms such as FolkRank, have the potential to leverage complex relationships in the data, corresponding to multiple hops in the graph. We present an in-depth analysis and evaluation of graph models for social tagging data and propose novel adaptations and extensions of FolkRank to improve tag recommendations. We highlight implicit assumptions made by the widely used folksonomy model, and propose an alternative and more accurate graph-representation of the data.
Our extensions of FolkRank address the new item problem by incorporating content data into the algorithm, and significantly improve prediction results on unpruned datasets. Our adaptations address issues in the iterative weight spreading calculation that potentially hinder FolkRank's ability to leverage the deep graph as an information source. Moreover, we evaluate the benefit of considering each deeper level of the graph, and present important insights regarding the characteristics of social tagging data in general. Our results suggest that the base assumption made by conventional weight propagation methods, that closeness in the graph always implies a positive relationship, does not hold for the social tagging domain.
\end{abstract}





\begin{bottomstuff}
Author's addresses: N. Landia (N.Landia@warwick.ac.uk), S. S. Anand (Sarabjot.Singh@gmail.com) {and} N. Griffiths (Nathan.Griffiths@warwick.ac.uk), Department of Computer Science, University of Warwick, Coventry CV4 7AL, UK; 
S. Doerfel (Doerfel@cs.uni-kassel.de), Faculty of Electrical Engineering and Computer Science, University of Kassel, Wilhelmsh\"{o}her Allee 73, 34121 Kassel, Germany;
R. J\"{a}schke (Jaeschke@l3s.de), L3S Research Center, Appelstrasse 9a, 30167 Hannover, Germany; 
A. Hotho (Hotho@informatik.uni-wuerzburg.de), Department of Computer Science, University of W\"{u}rzburg, Am Hubland, W\"{u}rzburg, Germany.
\end{bottomstuff}

\maketitle

\pagebreak
\section{Introduction}
Tagging is a popular methodology for many user-driven document organisation applications such as social bookmarking and publication sharing websites. On websites such as CiteULike\footnote{http://www.citeulike.org/}, Delicious\footnote{http://delicious.com/} and BibSonomy\footnote{http://www.bibsonomy.org/}, tags provide an unstructured organization method where each user has the liberty of choosing or making up any string of characters to be used as a tag for a document. The automatic generation of tag recommendations aids the social tagging process by reducing the effort required by the user, and making them aware of which existing tags are relevant to the document they are tagging. Tag recommenders encourage the use of the system and lead to a  more homogeneous document organisation overall. The task of tag recommendation is to automatically suggest a set of tags to a user for a document that he is in the process of tagging.

The data contained in social tagging systems is often described as a folksonomy \cite{Jaeschke2007}. A folksonomy is a tuple $(U, D,$ $T, A)$ where $U$ is the set of users, $D$ is the set of documents, $T$ is the set of tags and $A \subseteq U \times D \times T$ is the set of tag assignments. A tag assignment $a \in A$ is a triplet $(u, d, t)$ and indicates that user $u$ has assigned tag $t$ to document $d$. Thus a folksonomy can be modelled as a hyper-graph with the adjacency tensor given by a 3-dimensional binary matrix $F = [f_{i,j,k}]_{|U| \times |D| \times |T|}$ where each entry $f_{i,j,k} \in \{0,1\}$ specifies whether or not user $u_i$ tagged document $d_j$ with tag $t_k$. A post in the tagging data consists of a set of tags assigned by a user to a document. Posts themselves are only captured implicitly in the folksonomy model. The set of posts can be described as $P \subseteq U \times D \times 2^T$ where each post $p \in P$ is a triplet $(u_i, d_j, T_{ij})$ consisting of a user $u_i \in U$, a document $d_j \in D$, and a set of tags $T_{ij} \subseteq T$. We use the notation $(u_q, d_q, \emptyset)$ for query posts, where $u_q$ is the query user, $d_q$ is the query document and the set of tags is unknown and to be predicted, denoted by the empty set $\emptyset$.

Tag recommendation algorithms are often evaluated on post-core datasets which are subsets of the full tagging data. In a post-core at level $n$, only posts are included where the user, document and all tags appear in at least $n$ posts. This provides for a denser dataset and also minimises the chance that new items (users, documents or tags) will be encountered in the test data. However, post-cores only make up a small fraction of the full real-world data and the majority of posts are not included in this subset. Full tagging datasets have a large proportion of new/unseen documents. In order to recommend tags for these new documents, approaches are required which model documents not only based on the tags assigned to them in the past (if any), but also their content. While the inclusion of content data has been applied to keyword extraction and hybrid tag recommenders \cite{Lipczak2009,Lipczak2011}, graph-based approaches which model the full folksonomy graph as well as taking content data into account have not been widely explored.

\subsection{Contributions}
In this paper we present an in-depth analysis of the folksonomy graph model, propose novel adaptations and extensions to FolkRank \cite{Hotho2006}, and evaluate our hypotheses on four datasets from popular social tagging websites. We propose two alternative extensions to include content data into FolkRank's recommendation process and evaluate the predictive value of different content sources as well as varying amounts of content data. Our extensions make FolkRank successfully applicable to full unpruned datasets and address the new document problem in tag recommendation. In our extensive examination of the folksonomy graph model we highlight information that is lost and implicit assumptions that are made by the model, and propose a novel graph structure which captures the tagging data more accurately. Furthermore, we conduct an in-depth analysis of FolkRank's iterative weight spreading algorithm and address issues that exist therein. The outcome of this analysis is a novel weight spreading method which is much less computationally expensive while having a comparable recommendation accuracy to the iterative approach. Finally, we provide an extensive theoretical discussion as well as practical evaluation of the value of exploring the deep folksonomy graph. We evaluate whether the potential benefit of considering the information contained in deeper levels of the graph is worth the added computational expense and present important insights regarding the applicability of deep graph-based methods to tagging data in general. In summary, our main contributions are:

\begin{itemize}
\item Content-aware extensions of FolkRank which model the entire folksonomy as well as taking content data into account.
\item An improved graph data model which more accurately captures the tagging data.
\item A less expensive but equally accurate weight spreading method for graph-based tag recommendation.
\item An in-depth theoretical discussion as well as practical evaluation of the value of exploring the deeper folksonomy graph for tag recommendation, and of the applicability of graph-based methods to the domain of social tagging in general.
\end{itemize}

\section{Related Work}
\label{sec:related_work}
\begin{figure}
\begin{center}
\includegraphics[scale=0.45]{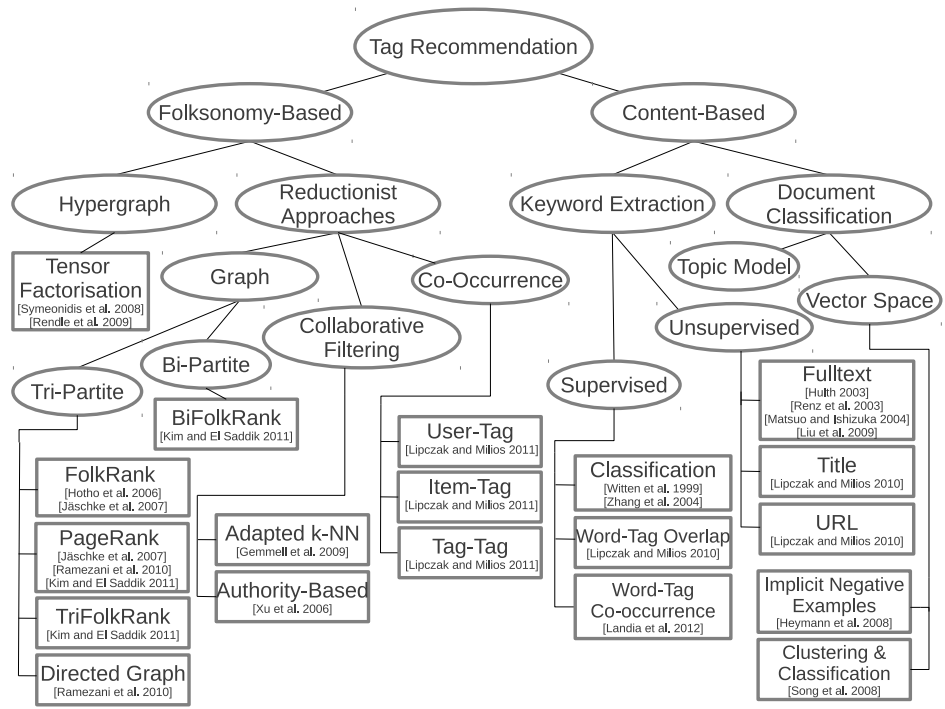}
\caption{Overview of Tag Recommendation Approaches}
\label{fig:approaches_overview}
\end{center}
\end{figure}

Existing tag recommendation solutions can be categorised into approaches which model and analyse the folksonomy in order to come up with recommendations, and content-based approaches where the textual content and/or meta-data of documents is considered. We give our view of the tag recommendation landscape in Figure \ref{fig:approaches_overview}. Methodologies relying on the folksonomy data include Hypergraph \cite{Symeonidis2008,Rendle2009}, Graph \cite{Jaeschke2007,Ramezani2010,Kim2011}, collaborative filtering \cite{Gemmell2009,Xu2006} and simple co-occurrence \cite{Lipczak2011} approaches. While hypergraph approaches try to capture and analyse all characteristics of the folksonomy in their models, graph-based and collaborative filtering approaches can be described as reductionist methods since they reduce the 3-dimensional folksonomy data to one or several 2-dimensional projections. A major difference between graph-based, collaborative filtering and co-occurrence approaches is that co-occurrence methods only consider the immediate neighbourhood of the query, corresponding to one hop in the folksonomy graph. Collaborative filtering considers connections one level deeper into the folksonomy, for example comparing the query user to similar users based on their overlap in document sets or tag vocabulary. Graph-based approaches have the possibility to explore the graph further and include information contained in the deep folksonomy into the recommendation process.

In content-based approaches, the textual content of the documents is used for either tag extraction and expansion \cite{Lipczak2009,Lipczak2011}, word-tag co-occurrence \cite{Landia2012}, or with document classification techniques \cite{Song2008}. Important aspects of content based approaches are the content source and the document representation used. Experiments have shown that the most informative words generally appear in the title and URL \cite{Lipczak2010b}, and the document text \cite{Heymann2008}. For structured text documents such as HTML, further sources such as anchors, links and paragraphs are available \cite{Zhang2004}. Heymann et al.\ carried out experiments on HTML pages \cite{Heymann2008}, comparing the value of page text, anchor text and text of surrounding hosts for tag prediction. They concluded that out of the three, the document text was most informative and anchor text was more informative than surrounding hosts. The document representation in content-based approaches is usually a bag-of-words. There are many different methods of determining the importance score of each word to the document, most of which include a Tf-Idf score in the calculation. \citeN{Liu2009} use Tf-Idf scores with part-of-speech analysis, word clustering and a sentence importance score; \citeN{Hulth2003} combines Tf-Idf scores and lexical tools; and \citeN{Renz2003} calculate the Tf-Idf scores on small word parts (quad-grams consisting of 4 letters) instead of whole words to overcome the stemming problem. In \cite{Witten1999}, Tf-Idf scores as well as the position of the first appearance of a word in the document are used. Alternatively to Tf-Idf, \citeN{Matsuo2004} use word frequency and word-word co-occurence to calculate scores for words given only single documents rather than a document collection.

In \cite{Lipczak2009}, Lipczak et al.\ present their hybrid tag recommender which won the content-based tag recommendation task of the ECML PKDD Discovery Challenge 2009 \cite{proceedings_dc09}. The part of the hybrid most comparable to our content-based approaches, for which Lipczak et al.\ give individual results, is ``user profile $\times$ resource related'', which is a combination of two tag recommendation sets: past tags of the query user and tags related to the content of the query document. The only source of content data in their approach is the document title. In order to generate the content-based tag recommendations, the words in the title of the query document which have been used as tags before in the training data are first extracted (word-tag overlap). The tag recommendation set is then expanded based on tag-tag co-occurence. Due to the initial filtering, the content words considered only include words which also appear as tags in the training data. The differences between our approaches for including content data and the content-related part of the hybrid in \cite{Lipczak2009} are the content sources and the document representation used. We consider all content words (after stopword removal) in our document representation, not only words that are tags, and base our predictions on word-tag co-occurence as well as utilising a content-based word importance measure (Tf-Idf). Moreover, we consider and evaluate two different content sources in our approaches: document title and fulltext content. The most successful part of the hybrid presented in \cite{Lipczak2009} is the ``Title Recommender'', which achieves better results than ``user profile $\times$ resource related''. It recommends words from the query document's title directly, choosing the words which have been observed to have a high global overlap in being a title word as well as a tag for documents. It is worth noting that these are words where the word string is equal to the tag string, which is not the case in our word-tag co-occurence approach. Nevertheless, it is a simple and very successful tag recommendation strategy producing a limited number of tags with high precision.

\subsection{FolkRank}
\label{sec:folkrank}
FolkRank \cite{Hotho2006,Jaeschke2007} is a graph-based ranking algorithm which is modelled based on Google's PageRank \cite{Brin98}. Similarly to PageRank, the key idea of FolkRank is that a document which is tagged by important users with important tags becomes important itself. The same holds symmetrically for users and tags. Users, documents and tags are represented as nodes $n \in N$ in an undirected tri-partite graph $G=(N,E)$, where all co-occurrences of users and documents, users and tags, and documents and tags are edges $e \in E$ between the corresponding nodes. The weight of the edge between two nodes depends on the number of their co-occurrences, given as the number of tag assignments that both nodes appear in. For example if a user $u$ has used a tag $t$ for two documents, there would be two tag assignments $(u, d_1, t)$ and $(u, d_2, t)$ in the folksonomy, and in $G$ the weight of the edge between the two nodes representing $u$ and $t$ would be set equal to two.

The importance or rank of each node in FolkRank is calculated by an iterative weight-spreading algorithm, in a similar fashion to PageRank. The weights of all nodes are given in the weight vector $\vec w$ which has one entry per node and is computed by the weight spreading function
\[\vec w \leftarrow (1-d)A \vec w + d \vec p\]
where $A$ is the row-stochastic version of the adjacency matrix of graph $G$, $\vec p$ is the preference vector, and the dampening factor $0 < d \leq 1$ determines the influence of $\vec p$. The preference vector $\vec p$ is used as a means to personalise the recommendations to the query user and document, and to achieve that goal is set to give the nodes representing the query user and document in the graph a higher preference weight compared to other nodes. The dampening factor $d$ sets the balance between personal preference and global importance when calculating the node weights. After constructing the folksonomy graph, the tag ranking procedure with FolkRank is as follows for each test post.
\begin{enumerate}
\item Initialise each node in the graph with a random starting weight so that the total sum of node weights in the graph is equal to a predefined parameter $TW$.
\item Set the preference vector giving the query user and document a higher weight than other nodes in the graph, and so that the sum of weights in the preference vector is equal to the total weight in the graph $TW$.
\item Perform iterative weight spreading until node weights converge. The end condition is that the sum of absolute change in node weights during one iteration is smaller than a predefined fraction of the total weight $TW$. 
\item Select the nodes which represent tags and rank them by node weight, where the tag node with the highest weight is given the best ranking.
\end{enumerate}
When setting the weights of the preference vector it is important that the sum of preference weights is equal to the total sum of node weights in the graph. This ensures that the total weight in the graph stays constant over weight-spreading iterations; that no factors other than parameter $d$ impact the amount of personalisation; and that the end condition of iterative spreading will work as intended. FolkRank can generate a global non-personalised ranking and a personalised ranking of all nodes in the graph, depending on the values set in the preference vector. For a global ranking, the entries in $\vec p$ for all nodes are set to the same value. In order to generate personalised recommendations for a query post $(u_q, d_q, \emptyset)$, $\vec p$ is set so that higher preference weights are given to the query user $u_q$ and query document $d_q$, compared to other nodes in the graph which are set to have a uniformly small (non-zero) preference weight \cite{Hotho2006}. The original FolkRank utilises a differential approach \cite{Jaeschke2007} to calculate the final tag scores for a query post. For each tag in the graph, the weight of the tag in the global ranking is subtracted from the weight of the tag in the query-personalised ranking to give the final prediction score. Although this approach produces better results than using the personalised ranking alone, a simpler alternative strategy is discussed in \cite{Kim2011}. Kim and El Saddik conclude that setting the preference weights of all non-query nodes in the personalised preference vector to zero instead of the uniformly small values results in an equivalent ranking to the differential approach.

An unexplored question in FolkRank is whether the same amount of preference weight should be given to the query user and query document. In the original FolkRank approach the balance of preference weight between the query user and document is not specified explicitly via a parameter but instead is determined implicitly by the ratio of user nodes to document nodes in the graph \cite{Jaeschke2007}. In general this leads to the query document receiving more preference weight than the query user since there are usually more documents than users in a folksonomy, however it is determined entirely by the data. We introduce a parameter $b$ into our FolkRank-based approaches which allows us to control how the total preference weight is distributed between the query user and document. The preference weights of the query user $u_q$ and query document $d_q$ are then given by
\[pw(u_q)=b * PW\] 
\[pw(d_q)= (1-b) * PW\] 
where $0 \leq b \leq 1$, $PW$ is the total preference weight and $PW = TW$ since we set the total preference weight equal to the total (starting) weight in the graph. If we set $b=\frac{|U|}{|U|+|D|}$, where $U$ is the set of users and $D$ is the set of documents, then we have the equivalent strategy to that used in the original FolkRank.

\section{FolkRank Adaptations}
\label{sec:folkrank_adaptations}
We conduct an in-depth analysis of the inner workings of FolkRank, highlight issues which might reduce tag recommendation accuracy and propose novel adaptations to overcome these. We first examine the folksonomy graph model used in FolkRank and the implicit assumptions made by the edge weights in the graph, and propose an alternative model which we call Post Graph. The second part of our analysis concerns itself with the weight-spreading iterations of FolkRank and the utilisation of information contained in the deep graph. We also address and propose how to overcome the high complexity and runtime of the algorithm.

\subsection{Graph Model and Edge Weights}
\label{sec:graph_construction}
The first issue we examine is the graph structure of the folksonomy model and the problem of setting the edge weights. FolkRank uses a tri-partite graph of the folksonomy consisting of user, document and tag nodes. Due to the fact that a post can contain a variable number of tags and since the post-membership information of tag nodes is not included explicitly in the folksonomy model, the user and document nodes in the graph can be connected to a variable number of tag nodes. The variable number of tags per post affects the outcome of weight spreading since in each spreading action the weight that is passed to each connected node depends on the total number and weight of edges of the active node. The difficulty is then setting the edge weights in the graph, where each alternative method of doing so makes different assumptions. In the following paragraphs we explore alternative graph construction methods and the implicit assumptions they make. We later evaluate each of these methods in Section \ref{sec:evaluation_and_results}. An assumption which holds in all alternatives is that the co-occurrence of users and tags, as well as documents and tags, should influence edge weights. The weight of a user-tag edge should be higher if the user has used the tag multiple times to tag multiple documents. Similarly, if the same tag has been assigned to a document multiple times, so by multiple different users, the document-tag edge should be given a higher weight. In contrast, for user-document relationships the tagging data only provides one distinct co-occurrence since a user can only tag each document once (with a set of tags).

\paragraph{Folksonomy Graph}

\begin{figure}
\begin{center}
\includegraphics[scale=0.4]{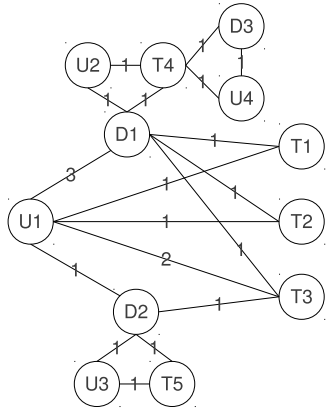}
\end{center}
\caption{Folksonomy Graph}
\label{fig:folkrank_alt1}
\end{figure}

The folksonomy graph structure and edge weighting methodology used in original FolkRank is given in Figure \ref{fig:folkrank_alt1}. The weight between user and document nodes is set according to the number of tag assignments and thus the number of tags in the post (see edge weights $u_1$-$d_1$ vs $u_1$-$d_2$). This means that within the context of a post, all types of nodes (user, document, and all tags together) get the same amount of total weight. In the context of post $(u_1, d_1, [t_1, t_2, t_3])$ only, ignoring the influence of post $(u_1, d_2, [t_3])$, the weight of the edge $u_1$-$d_1$ is the same as the sum of edge weights $u_1$-$t_1$, $u_1$-$t_2$ and $u_1$-$t_3$, which is 3. However, another consequence of this graph construction method is that, if we spread weight from $u_1$, then $d_1$ would get a higher weight than $d_2$, and subsequently, the tags connected to $d_1$, in this case $t_4$, would get a higher weight than the tags connected to $d_2$, namely $t_5$. The implicit assumption made by this model is that documents to which a user has assigned many tags are more representative of the user's interest. Another assumption is made with regard to the number of tags in a post. If we had a query post $(u_1, d_3, \emptyset)$, the fraction of weight spread from $u_1$ to $t_3$, which is the user's most used tag, would be ${2}/{8}$ (times the dampening factor). However, the query document $d_3$ would spread ${1}/{2}$ of its weight to $t_4$. Assuming both the query user and query document have the same preference weight, $t_4$ would thus be ranked higher than $t_3$ even though $t_3$ has been used by the user multiple times and $t_4$ has only been assigned to $d_3$ once. The assumption which leads to this outcome is that if a post has multiple tags then each of the tags is proportionally less important to the user and document of the post.
In summary, as a consequence of the graph model, the following implicit assumptions are made when spreading weight in the folksonomy graph.
\begin{itemize}
\item Within the context of a post, all types of nodes (user, document, tag) have the same amount of relevance summed by node type.
\item The weight of the user-document relationship depends on the number of tags in the respective post. The more tags a user has assigned to the document, the stronger the user-document connection.
\item Each tag in a post is proportionally less important to the user and document if the post contains multiple tags.
\end{itemize}

\paragraph{Folksonomy Adapted Graph}

\begin{figure}
\begin{center}
\includegraphics[scale=0.4]{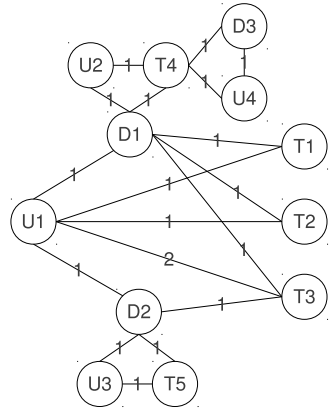}
\end{center}
\caption{Folksonomy Adapted Graph}
\label{fig:folkrank_alt2}
\end{figure}

We propose an alternative edge weighting method for the folksonomy graph, illustrated in Figure \ref{fig:folkrank_alt2}, which we refer to in our experiments as the Adapted Graph (AG). The difference to the original folksonomy methodology is that we always keep user-document edges at a weight of 1 regardless of the number of tags in the post. By spreading weight in the Adapted Graph, the following assumptions are made.
\begin{itemize}
\item Within the context of a post, all tag nodes together are more important than user or document nodes.
\item The weight of the user-document relationship is independent of of the number of tags in the respective post. All edges connecting users to documents have the same weight.
\item Each tag in a post is proportionally less important to the user and document if the post contains multiple tags.
\end{itemize}

\paragraph{Post Graph}
\label{sec:postrank}

\begin{figure}
\begin{center}
\includegraphics[scale=0.4]{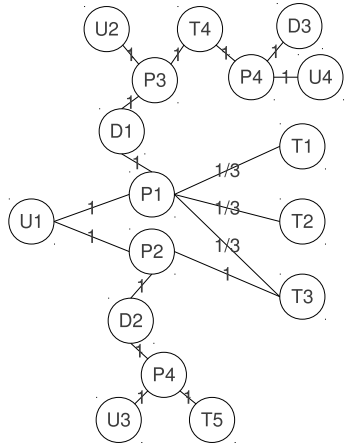}
\end{center}
\caption{Post Graph}
\label{fig:postrank1}
\end{figure}

Since both aforementioned graph construction methods do not explicitly include the post-membership information of tag nodes, we believe that they produce an inaccurate model of the social tagging data, and propose a structurally different graph model which we call Post Graph (PG). The Post Graph model includes an additional type of node representing posts themselves into the graph. Figure \ref{fig:postrank1} shows the Post Graph for the same data as the previous folksonomy graph and Adapted Graph models. The user, document and tag nodes are only connected to post nodes instead of being directly connected to each other, and we set the weight of post-tag edges so that the edge weights to all tags of a single post sum to 1. This makes the strength of the user-document relationships independent of the number of tags in the post, as well as ensuring that the same amount of total weight is spread to all types of nodes in the context of a post. To address the assumption that having multiple tags in a post implies less importance for each of them we evaluate an alternative method of retrieving tag scores from the graph. Instead of directly retrieving tag scores as the weight of tag nodes, we retrieve the weight of post nodes instead, and in a second step calculate the tag scores by summing up for each tag the weight of the post nodes that the tag is related to (ignoring the total number of tags in each post).

\begin{itemize}
\item Within the context of a post, all types of nodes (user, document, tag) have the same amount of relevance summed by node type.
\item The weight of the user-document relationship is independent of of the number of tags in the respective post. All edges connecting users to documents have the same weight.
\item By retrieving each tag's score as the sum of weights of post nodes it is connected to, the importance of each tag is independent of the total number of tags in its post.
\end{itemize}

\subsection{Weight Spreading and Value of Deep Graph}
\label{sec:weight_spreading}
\begin{figure}
\begin{center}
\includegraphics[scale=0.5]{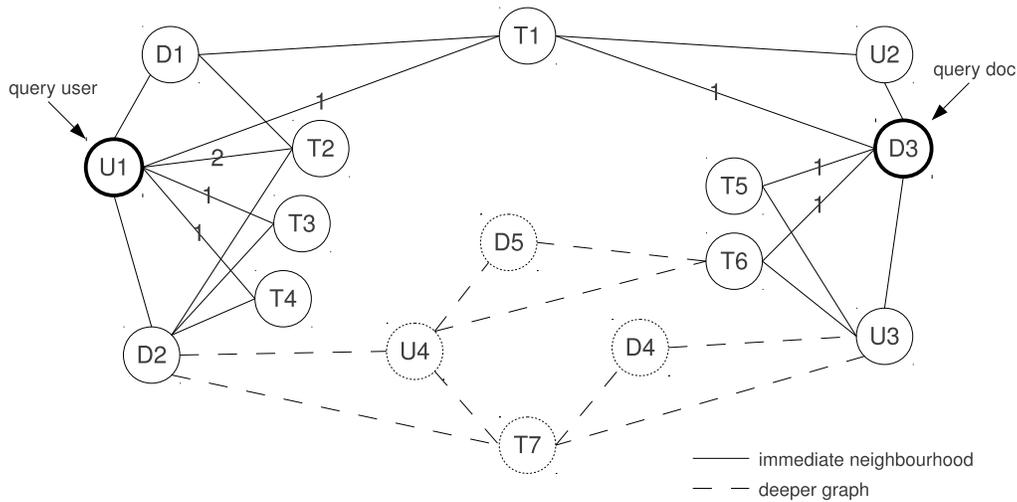}
\end{center}
\caption{FolkRank Utilisation of Deep Graph}
\label{fig:folkrank_shallow_vs_deep}
\end{figure}

\subsubsection{Discussion}
\label{sec:weight_spreading_discussion}
FolkRank's iterative weight spreading algorithm has two potential advantages over approaches which only utilise the immediate neighbourhood of the query nodes, such as simple co-occurrence methods. Firstly, by initialising all nodes with random starting weights, the general importance of tags is taken into account when generating recommendations. This can also be described as authority-based popularity, due to the characteristic that important user or document nodes will provide more weight to their connected tags. 
Secondly, the weight spreading algorithm considers the information contained in the deep graph. In Figure \ref{fig:folkrank_shallow_vs_deep} we illustrate how FolkRank utilises the deeper graph beyond the immediate neighbourhood of the query user and document. User $u_1$ and document $d_3$ make up the query post, the immediate neighbourhood of the query nodes is shown in solid lines, the deeper graph is shown in dashed lines, and the weights of all edges which are not explicitly labelled are set to 1. In approaches considering only the immediate neighbourhood of $u_1$ and $d_3$, the candidate tag set for this query post would consist of tags $t_1$, $t_2$, $t_3$, $t_4$, $t_5$, and $t_6$. A co-occurrence approach, such as our combination of user-related and document-related tags \cite{Landia2012}, would rank $t_1$ as the best recommendation as it is related to both $u_1$ and $d_3$, followed by $t_2$ as the second best since it has a relatively strong relationship with $u_1$. However, when trying to rank $t_5$ and $t_6$, both of these tags would have the same prediction score and the algorithm would not have sufficient information to decide which of them should precede the other in the ranking. In the final tag predictions, the ordering of $t_5$ and $t_6$ would be random. By utilising the deeper graph, FolkRank's iterative weight spreading algorithm has the ability to provide a definitive ranking of $t_5$ and $t_6$ by trying to deduce which of them is more important to the query nodes. It would spread weight along the path $u_1 \rightarrow d_2 \rightarrow u_4 \rightarrow t_6$ and thus $t_6$ would be ranked higher than $t_5$. The other method that FolkRank has for breaking ties and re-ranking tags which would otherwise have equal prediction scores are the general importance weights. Additionally, FolkRank also spreads weight to tag $t_7$ found in the deeper graph and includes it in the candidate set, whereas $t_7$ would be omitted by approaches which only recommend tags co-occurring with the query user or document.

It seems intuitive from the graph structure and from literature applying graph-based approaches to non-folksonomy data that $t_7$ should receive some weight and be included in the candidate tag set, and that $t_6$ is more related to the query and thus should be ranked higher than $t_5$. However, the value of following this computationally expensive strategy and considering the connections of the deep folksonomy graph has not yet been directly evaluated. As there are other factors that impact the weight spreading calculation of FolkRank, which we explore in the following paragraphs, it has not yet been established that considering the deep graph provides an increase to tag recommendation accuracy. As part of the theoretical discussion, the opposite argument to FolkRank's assumption about the predictive indications of the deep graph could also be made. Considering the query user $u_1$, if we make the somewhat weak assumption that $u_1$ is aware of the existence of tag $t_7$, then it would not make sense to spread weight to $t_7$. In the graph in Figure \ref{fig:folkrank_shallow_vs_deep}, user $u_1$ has tagged $d_2$ with the tags $t_2$, $t_3$ and $t_4$. A different user $u_4$ has also tagged the same document $d_2$ with tag $t_7$. If we assume that $u_1$ has, in his view, completely described $d_2$ with $t_2$, $t_3$ and $t_4$, this would suggest that $t_7$ was not required by the query user $u_1$ to describe $d_2$. One could thus argue that this was a conscious decision and $t_7$ might not be considered to be a good descriptor by $u_1$ in general. The weak points of this argument are the generalisation and the assumption of completeness. Rather than dismissing tag $t_7$ completely, $u_1$ might also think that $t_7$ is not appropriate for the documents he has tagged so far but generally a useful descriptor. More importantly, user $u_1$ might not be aware of tag $t_7$ at all. However, starting from the query document $d_3$, a similar and more convincing argument can be made with regard to $t_7$. Document $d_3$ has been tagged with $t_5$ and $t_6$ by user $u_3$, who has also tagged a different document $d_4$ with $t_7$. In this case the argument against assigning a higher weight to $t_7$ as a candidate tag for $d_3$ is much stronger. Since user $u_3$ has used $t_7$, it is guaranteed that he is aware of its existence. However, $u_3$ has explicitly not assigned $t_7$ to $d_3$, and is using the different tag sets of $[t_5, t_6]$ and $[t_7]$ to distinguish between documents $d_3$ and $d_4$. If any deduction is made about the relevance of $t_7$ to $d_3$, it should be that the graph indicates a negative relationship and the weight of $t_7$ with regard to $d_3$ should be reduced rather than increased. 

The counter-argument to utilising even longer paths, which leads to FolkRank's ranking of $t_6$ above $t_5$, is the highly personal tagging behaviour of users in (broad) folksonomies. FolkRank uses the path $u_1 \rightarrow d_2 \rightarrow u_4 \rightarrow t_6$ to deduce that $t_6$ is more relevant than $t_5$ to the query consisting of user $u_1$ and document $d_3$. However, this deduction is based on the fact that $t_6$ was used by a different user $u_4$ for a different document $d_5$, and the only link to the query is given by $u_4$ having tagged $d_2$ which has also been tagged by the query user $u_1$. The shared document $d_2$ is taken as an indication that $u_1$ and $u_4$ have similar interests and that $u_1$ should give some authority to all of the other opinions/tag assignments made by $u_4$. \citeN{Wetzker2010} argue that tag assignments cannot be transferred as easily across users and provide evidence for the highly personalised tagging behaviour of users in broad folksonomies. They show that users who have tagged the same documents rarely assigned the same tags to these documents. Even though the users' areas of interest are similar due to the shared documents, only a small overlap can be observed in their tag vocabulary which indicates that the users' views of the documents are highly personal.

\subsubsection{Analysis of Iterative Weight Spreading in Folksonomies}
In the following paragraphs we analyse the iterative weight spreading method of FolkRank in detail and address issues which we believe to hinder or cascade its ability to effectively utilise the information contained in the deeper graph. An important preliminary observation about FolkRank is that the impact of each preference node on the final weights in the graph is independent of the influence of other preference nodes. Having multiple nodes with preference weight $>0$ in the preference vector of FolkRank is virtually equivalent to performing the weight spreading computation for each of the preference nodes separately, and then doing a linear combination of the resulting weight vectors to give the final ranking. As long as the end condition of the weight-spreading iterations is set sufficiently small, the only nodes which could end up with a different weight are the ones at the very bottom of the ranking. This method can be used to speed up the prediction time of FolkRank in a live tag recommendation scenario. For each user in the system, the tag scores can be pre-calculated offline and stored. The same can be done for each document. During the online tag recommendation phase, the pre-calculated tag scores for the query user and query document would then be retrieved and a weighted average of the scores would be taken per tag in order to quickly create tag recommendations. For the following discussion we assume that this method of performing a separate weight spreading computation for each of the preference nodes is used, so that each individual weight spreading run will have only one node in the preference vector.

\paragraph{Swash-Back Problem}
\begin{figure}
\begin{center}
\includegraphics[scale=0.4]{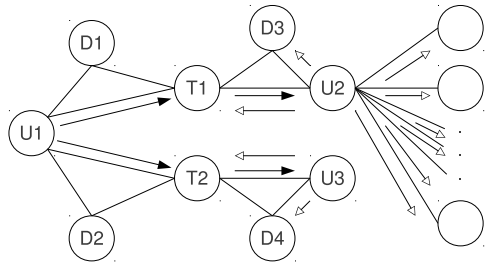}
\end{center}
\caption{FolkRank Swash-Back Problem}
\label{fig:folkrank_swashback}
\end{figure}

A problem of FolkRank as discussed in \cite{Jaeschke2007} is ``swash-back'' of weights. Since the graph is undirected, weight is spread from a node $n_1$ to a connected node $n_2$ in one iteration and then spread back from $n_2$ to $n_1$ in the next iteration. This means that the weight of $n_1$, the node from which the weight-spreading originates, is adjusted in the second iteration based on the (number of and weight of) edges of $n_2$, which does not seem intuitive and is not desirable. We illustrate the consequences of this in Figure \ref{fig:folkrank_swashback}. User $u_1$ has tagged documents $d_1$ and $d_2$ with tags $t_1$ and $t_2$ respectively. Tag $t_1$ is also used by a very active user $u_2$ who is connected to a large number of other nodes, where $t_2$ is also used by user $u_3$ who has only one tag assignment. If we want to recommend tags for $u_1$ and activate that node in the graph, $t_1$ and $t_2$ would get the same weight in the first iteration. In the second iteration $t_1$ spreads weight to $u_2$, and $t_2$ spreads weight to $u_3$ (as well as all of their other connected nodes), where the weight received by $u_2$ and $u_3$ is equal. The third iteration is when the swash-back with regard to $t_1$ and $t_2$ occurs, denoted by the empty arrows. Tag $t_2$ gets half of the weight of $u_3$ (times the dampening factor) back since $u_3$ is connected to two nodes. However, $u_2$ is connected to many other nodes, and so the weight spread back from $u_2$ to $t_1$ would be much less than the weight spread back from $u_3$ to $t_2$. In the final tag predictions for user $u_1$, tag $t_2$ would have a higher score than $t_1$ due to the behaviour of users $u_2$ and $u_3$. This is contrary to our intuition that the weights should be equal up to this point since the query user $u_1$ has used both tags with equal frequency in the past. In the final ranking, when the node weights of the query user are combined with the node weights produced by the query document, the weights of $t_1$ and $t_2$ are expected to change to reflect the influence of the deeper graph. However, in this weight spreading operation for the query user only, the only source of preference weight is $u_1$. The change in weights due to swash-back might outweigh the later influence of other preference nodes and prevent FolkRank from utilising the information contained in the deeper graph.

\paragraph{Triangle-Spreading Problem}
\begin{figure}
\begin{center}
\includegraphics[scale=0.4]{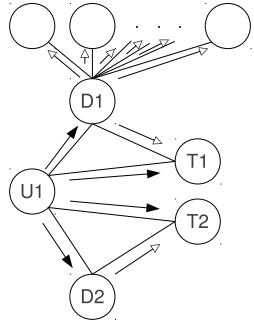}
\end{center}
\caption{FolkRank Triangle Spreading Problem}
\label{fig:folkrank_triangle_spreading}
\end{figure}
Another issue, which we refer to as triangle-spreading of weights, is illustrated in Figure \ref{fig:folkrank_triangle_spreading}. User $u_1$ has tagged document $d_1$ and $d_2$ with tags $t_1$ and $t_2$ respectively. Document $d_1$ is a popular document tagged by many other users, whereas $d_2$ has only been tagged by $u_1$. If we activate $u_1$ in order to recommend tags for this user, tags $t_1$ and $t_2$ would get the same weight in the first iteration. In the second iteration, $d_2$ would spread half of its weight to $t_2$ (times the dampening factor), however, $d_1$ would spread less of its weight to $t_1$ since $d_1$ is also connected to several other nodes. This would mean that in the tag weights for query user $u_1$, tag $t_2$ would get a higher weight than $t_1$ even though the user has used both tags with equal frequency. A similar problem would arise with regard to the weight of documents $d_1$ and $d_2$ if one of the tags was very popular. Due to graph being undirected and the folksonomy consisting of triplet relationships (user, document, tag), if two nodes $n_1$ and $n_2$ are connected, there is always at least one indirect path from $n_1$ to $n_2$ via a third node $n_3$. The weight spread from $n_1$ to $n_2$ over the indirect path via $n_3$ is influenced by the (number of and weight of) edges of $n_3$. This is undesirable since the weight of the direct edge from $n_1$ to $n_2$ already completely describes the relationship between $n_1$ and $n_2$. Moreover, the influence of the triangle-spreading is likely to cascade the effect of the deeper graph on final tag weights, since the indirect path along which the undesired spread happens has a length of only two hops.

\subsubsection{PathRank}
\label{sec:pathrank}
In order to address the swash-back and triangle spreading problems we present our adapted weight-spreading approach for undirected folksonomy graphs, which we call PathRank. Rather than doing iterative weight spreading, PathRank assigns scores to each node in the graph based on the shortest path(s) from the preference nodes. The weight spreading computation works in a similar manner to a breadth-first search, where edges which were already explored in previous iterations are not re-visited. PathRank is akin to spreading activation which is usually applied to directed graphs, and where nodes also spread their weight only once. However, PathRank is used on the un-directed folksonomy graph and gives the edges a personalised direction starting from the query nodes, where the edge direction can be different for each query. PathRank can be described as activation-directed weight spreading. In contrast to the original iterative weight spreading approach of FolkRank, we set the initial weight of all nodes in the graph to zero instead of initialising nodes with random starting weights. PathRank thus only uses personalised weights, originating from the preference nodes, and there are no general importance weights in the graph (which makes the dampening factor parameter obsolete). We compute node weights separately for each preference node and then take a weighted average of the resulting weights for each node in the graph to produce the final node weights, taking into account all preference nodes. Because of the separate calculation per preference node and the setting of all starting weights in the graph to zero, each individual weight-spreading calculation has only one node, the preference node, from which all of the weight in the graph originates. The swash-back and triangle spreading of weights can then be prevented by adapting the iterative weight spreading algorithm with a simple rule: \textit{If the weight of a node has been updated in a previous iteration (i.e.\ is not equal to zero), then do not re-calculate the node's weight in subsequent iterations}. Thus, the PathRank weight spreading algorithm is in effect not an iterative update calculation like in PageRank/FolkRank, but rather assigns a weight to each node $n_i$ based on the edges of the shortest path(s) from the each of the preference nodes $n_p$ to $n_i$. The parameter $pl$ specifies the maximum path length from the preference nodes to be explored by PathRank. The end condition of PathRank weight spreading is that either the maximum path length $pl$ has been reached, or that all nodes in the graph have been explored and assigned a weight greater than zero.

The benefits of PathRank are that the problems of swash-back and triangle-spreading of weights are removed, which allows the algorithm to fully utilise the information contained in the deeper graph. Since there are no general importance weights, these also cannot interfere with and cascade the influence of the weight spread through the deeper graph. Intuitively we would assume that weights spread from preference nodes through the deeper graph would result in a better re-ranking of the tag nodes in comparison to using general importance weights, since the general importance of nodes is not personalised and constant across all query posts. Setting different values for the maximum path length $pl$ to be explored allows for a direct evaluation of the value of including the deeper graph in the recommendation process. In our evaluation in Section \ref{sec:tuning_pl} we address the question of how much value there is in exploring each step deeper into the graph when calculating tag predictions. 

Regarding runtime, as long as we only have one preference node, the complexity of weight spreading is greatly reduced in PathRank compared to FolkRank, since once a node's score is set it does not need to be re-calculated in every subsequent iteration. If we take the same graph, let $i$ denote the total number of iterations and $n$ denote the number of edges in the graph, FolkRank's iterative weight spreading has a complexity of $\mathcal{O}(2n \cdot i)$. In each iteration, weight is spread in both directions along each edge, partly because the nodes are initialised with random starting weights. PathRank has a worst-case complexity of $\mathcal{O}(n)$ if the weight spreading is performed until all nodes in the graph are explored. Weight is only spread once along each edge in one direction. However, in the case that there are several preference nodes, PathRank needs a separate weight-spreading calculation for each of them, meaning the complexity would be $\mathcal{O}(n \cdot p)$ where $p$ is the number of nodes with weight $>0$ in the preference vector, whereas the runtime of FolkRanks's iterative algorithm would not change. For the expensive FolkRank algorithm to be applicable in practice, the individual tag scores per user and per document have to be pre-calculated offline, and then combined in the online recommendation phase to quickly generate predictions. In this scenario, where each of the pre-calculation runs has only one node in the preference vector, PathRank is guaranteed to outperform FolkRank regarding runtime. Moreover, by limiting the maximum path length via the parameter $pl$, the runtime can be further reduced. As we later show in the evaluation, $pl$ can be set to almost minimal values without a decrease in prediction accuracy.

\section{Extension with Content-Data}
Here we present our methods for extending FolkRank with content data. These content-aware recommenders include the textual content of documents in the recommendation process as well as utilising the full folksonomy graph. This allows us to relate new unseen documents to already tagged (different) documents in the system and make recommendations based on the tag assignments related to those known documents. We can thus overcome the new document problem and make the solely folksonomy-based recommenders applicable to full real-world datasets. For test posts where the query user is new (as well), we have to default to the most popular tags found to be related to the content of the query document and cannot personalise these to the user, which is acceptable since the user does not have a tagging profile yet. In the following sections we first describe the document content model we use and then present our content-aware graph recommenders.

\subsection{Document Model}
\label{sec:doc_model}
For including data from the content of documents in the tag recommendation algorithms, we consider two sources of content words: document title and fulltext content. We convert all words to lower-case, remove stop-words as well as all words which have a length of less than 3 or more than 20 characters, and use the remaining words without stemming.
Each document is represented by a bag-of-words vector of content words with Tf-Idf scores for each word. Tf-Idf stands for \textit{Term frequency-Inverse document frequency} and we compute it as
\[{\text{Tf-Idf}(w_l,d_j) = {\frac{\text{tc}(w_l, d_j)}{|d_j|}}* \log_2{\frac{|D|}{\text{dc}(w_l, D)}}}\]
where $D$ is the set of all documents, $\text{tc}(w_l, d_j)$ is the term count equal the number of occurrences of word $w_l$ in document $d_j$, and $\text{dc}(w_l, D)$ is the document count equal to the number of documents in the collection containing word $w$. We normalise the Tf-Idf scores to sum to 1 per document. 

A factor to consider is that the content data of websites can change over time. The title, content and meta-data of a website which is bookmarked can be updated and differ from one post to the next. This presents a problem, as well as additional data for analysis. The fulltext content of the bookmarked website itself is only available in the current version at the time of retrieval, however, the BibSonomy dataset provides different versions of metadata for a document at the time of each post. Where available, we concatenate the title variations of a document from all its posts and treat the resulting text string as the single document title. This makes the term count measure $\text{tc}(w_l, d_j)$ in our Tf-Idf calculation more powerful as words which persist over several title variations will end up with a higher score than words which only appear in a few of the variations.

\subsection{ContentFolkRank}
\label{sec:content_folkrank}
\begin{figure}
\begin{center}
\includegraphics[scale=0.4]{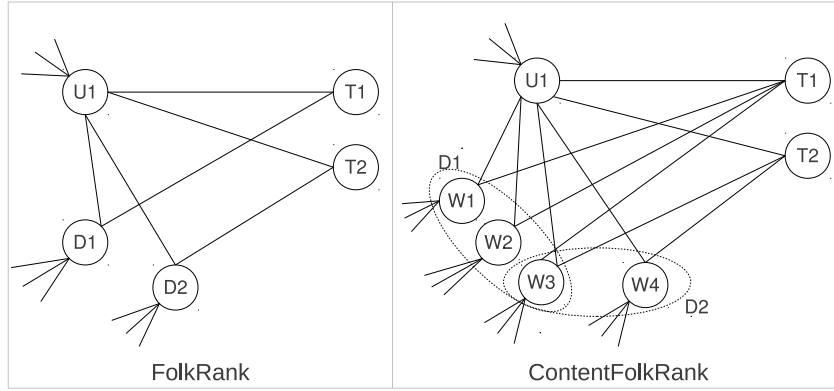}
\end{center}
\caption{FolkRank and ContentFolkRank}
\label{fig:folkrank_document_vs_word_nodes}
\end{figure}

ContentFolkRank (which we first presented in \cite{Landia2012}) includes the content of documents directly into the graph. We adapt the original folksonomy graph of FolkRank to model triplets $\mathit{(user, word,}$ $\mathit{tag)}$ instead of 
$\mathit{(user, document, tag)}$. Each tag assignment in the training data $\mathit{(u, d, t)}$ is converted to a set of tag assignments with words instead of documents $\{(u, w_1, t)$, $(u, w_2, t)$, $\dots$, $(u, w_k, t)\}$ where each of the words $w$ is in the content of $d$. Figure \ref{fig:folkrank_document_vs_word_nodes} shows the standard FolkRank graph with document nodes on the left, and the ContentFolkRank graph modelling the same data on the right, where document $d_1$ is represented by word nodes $w_1$, $w_2$ and $w_3$, and document $d_2$ is represented by word nodes $w_3$ and $w_4$.

In ContentFolkRank we use custom rules for setting the weights of different types of edges, namely user-word edges, word-tag edges and user-tag edges. To prevent the content length of a document, and thus the number of word nodes representing the document in the graph, from influencing the recommendation process, we utilise Tf-Idf scores when setting the edge weights. The Tf-Idf scores are normalised to sum to 1 per document. This provides suitable weights for the edges of word nodes representing the document in the graph and ensures that the number of words itself does not impact the generated recommendations. Additionally, the Tf-Idf scores provide appropriate weights for capturing the varying importance of different content words to the document, and have been shown to be beneficial in \cite{Landia2012}. Since several documents, in our example ${d_1}$ and ${d_2}$, tagged by the same user ${u_1}$ can contain the same word ${w_3}$, the weight of the edge between ${u_1}$ and ${w_3}$ is set to the sum of the normalised Tf-Idf scores of ${w_3}$ in ${d_1}$ and ${d_2}$. The same holds for edges between word and tag nodes. The final edge weights thus take the content importance of words as well as tagging co-occurence in the folksonomy into account. The weight of the edges between user and tag nodes are solely based on co-occurrence since only complete documents (and not individual words) can be tagged by users, and are set to the number of posts in which the user has used the tag.

The formulae for calculating the weights of the different types of edges are the following. For user-word edges, the edge weight is given by
\[{\text{edgeWeight}(u,w) = \sum_{d_j \in \text{Posts}(u,w)} \text{Tf-Idf}(w,d_j)}\]
where $\text{Posts}(u,w)$ is the set of posts by user $u$ where the document contained word $w$. Similarly, for word-tag edges we calculate the weight using
\[{\text{edgeWeight}(w,t) = \sum_{d_j \in \text{Posts}(w,t)} \text{Tf-Idf}(w,d_j)}\]
where $\text{Posts}(w,t)$ is the set of posts tagged with $t$ (by any user) where the document contained word $w$. For user-tag edges there is no need to include Tf-Idf scores as complete documents are tagged by users and words cannot be tagged on their own, so we use the formula
\[{\text{edgeWeight}(u,t) = \mid\text{Posts}(u,t)\mid}\]
where $\text{Posts}(u,t)$ is the set of posts where user $u$ used tag $t$.

The preference vector for each test post is given by ${(u_q, w_1, w_2, \dots, w_k)}$ where $u_q$ is the query user and each word $w$ is in the content of the query document $d_q$. The preference weight for each word $w$ is set proportional to its Tf-Idf score in $d_q$, and is given by
\[pw(w)= \text{Tf-Idf}(w, d_q) * (1-b) * PW \]
where the Tf-Idf weights are normalised to sum to 1 per document, $b$ is the parameter for setting the balance in preference weight between the query user and document, and $PW$ is the total preference weight. The preference weight of the query user is the same as before without content: $pw(u_q)=b * PW$.

\subsection{SimFolkRank}
\label{sec:SimFolkRank}
Our second approach of including content into the recommendation process is to utilise a content-based document similarity measure and include content information implicitly rather than introducing words directly into the graph. The graph model of SimFolkRank does not contain content data itself and documents are represented by document nodes, using either the original folksonomy graph (SimFolkRank) or the Post Graph model (SimFolkRank\_PG). However, for each test post, we construct the preference vector to include not only the query document (if it already exists in the graph) but also a predefined number of training documents most similar in content to the query document. In our experiments we evaluate the effects of including different numbers of most similar documents in the preference vector. The similarity between documents is calculated based on the words in either the title or the full text of the documents. The metric we use is cosine similarity of the bag-of-words document vectors with normalised Tf-Idf scores. Due to the problem that document content data can vary over time (as discussed in Section \ref{sec:doc_model}), it can be the case that a query document which also exists in the graph as a training document ends up with a low content similarity score with itself. To overcome this issue, we include an additional step where we set the similarity of query documents with themselves to 1, provided that they appear as training documents as well. Once the cosine similarity of a query document to each training document is calculated, we normalise the similarity scores to sum to 1 for the query document. This ensures that the number of similar training documents with cosine similarity greater than zero does not affect the recommendation process.

The preference weight of each training document $d$ included in the preference vector is a function of its similarity to the query document $d_q$ and is given by
\[pw(d) = \text{sim}(d, d_q) * (1-b) * PW \]
where $\text{sim}(d, d_q)$ is the content-based similarity between $d$ and $d_q$ normalised to sum to 1 over all documents $d$ similar to $d_q$, parameter $b$ determines the balance in preference weight between the query user and document, and $PW$ is the total preference weight. The preference weight of the query user is the same as before: $pw(u_q)=b * PW$. We apply and evaluate this approach of including content with iterative weight spreading (SimFolkRank, SimFolkRank\_PG) as well as our PathRank weight spreading algorithm (SimPathRank, SimPathrank\_PG).

\section{Experimental Setup}

\subsection{Datasets}
Our datasets consist of tagging data from the social bookmarking websites CiteULike\footnote{http://www.citeulike.org/}, Delicious\footnote{http://delicious.com/} and BibSonomy\footnote{http://www.bibsonomy.org/}, and additionally downloaded content data for our content-aware recommenders. Official snapshots of CiteULike and BibSonomy are available on their respective websites. We use the CiteULike 2012-05-01 snapshot, and the BibSonomy 2012-07-01 snapshot. The BibSonomy social bookmarking website and dataset is split into two separate sections: BibSonomy Bookmark which are website bookmarks and BibSonomy BibTeX which are publication bookmarks. We treat these two subsets of BibSonomy as separate datasets. Delicious does not provide snapshots of their data. Here we use a dataset that was obtained by crawling the Delicious website in 2005, the specifics of which are given in \cite{Hotho2006}.

Additionally, we downloaded all of the available pages from the URLs in the Delcious and BibSonomy Bookmark datasets, and all of the BibTeX entries for CiteULike. For our content-aware recommenders the two content data sources are the title and fulltext for websites, and the title and abstract for publications. Our Delicious crawl and the BibSonomy Bookmark and BibSonomy BibTeX snapshots provide the titles of documents. For CiteULike we extracted the titles from the downloaded BibTeX entries. The fulltext content for Delicious and BibSonomy Bookmark is the page text of the bookmarked websites, which we extracted from the downloaded pages. For CiteULike and BibSonomy BibTeX, where the bookmarked documents are publications, we use the abstracts from the BibTeX entries as the fulltext content.

\subsection{Pre-Processing}
We pre-processed all of the datasets by casting all tags to lower case, removing duplicate tag assignments that might occur as a result of this, and removing posts which have no tags. Additionally, for CiteULike there are some automatically generated tags which occur very frequently. In order to clean the dataset of these tags, we removed all tag assignments where the tag equals ``no-tag" or ``bibtex-import", or matches the regular expressions ``*file-import*" or ``*import-*". For BibSonomy BibTeX we removed all tag assignments where the tag is ``jabrefnokeywordassigned" or ``myown" since these occur with disproportional frequently and can single-handedly skew results.

\subsection{Evaluation Metrics}
\label{sec:evaluation_metrics}
We use recall@$N$, precision@$N$ and F1@$N$ as our success measures, where $N$ is the predefined number of tags to be recommended. Recall measures the ratio of correct recommendations to the number of true tags of a test post, whereas precision measures the ratio of correct to false recommendations made. Recall and precision are given by
\[{recall = \frac{\text{TP}}{\text{TP} + \text{FN}}}\]
\[{precision = \frac{\text{TP}}{\text{TP} + \text{FP}}}\]
where $\text{TP}$ (true positives) is the number of correct tags recommended, $\text{FP}$ (false positives) is the number of wrong recommendations and $\text{FN}$ (false negatives) is the number of true tags which were not recommended. F1 is a combination of recall and precision and is given by
\[{\text{F1} = \frac{2 * precision * recall}{precision + recall}}\]
Since we believe recall to be more important than precision in the context of tag recommendation, as long as $N$ is kept reasonably low ($<=$10), we use recall in the evaluation phase to identify the best recommenders and configurations. We then give recall as well as F1 for the final results.

\subsection{Training and Test Sets for Unpruned Tagging Data}
To construct a training and test set for the experiments on the full/unpruned tagging data, we use the following date-split approach for each of the datasets. 
The test set consist of all posts in the most recent two months of the data which provides us with a large enough test set size. The resulting numbers of test posts are 76,491 for CiteULike, 1.7M for Delicious, 9,506 for BibSonomy Bookmark and 2,843 for BibSonomy BibTeX. The training set is a sample of the data prior to the two test months. We use a sample and not the complete historical data for our training set since the FolkRank-type algorithms have a high computational complexity and expense. Note that we only apply sampling for the training dataset, while in the test set all posts made in the test time-frame are included.

The aim of our sampling methodology for the training set is to achieve a small enough sample size for our models to generate recommendations within a reasonable time while introducing as little bias into the models as possible.
To create the training sample we start by selecting the 12 months of data prior to the test months. Social tagging datasets have been shown to be time-sensitive where popular post topics as well as users' interests change over time, and we believe that posts which are older than a year from the test period provide less predictive data for generating recommendations. We then take a stratified sample of documents, where the stratification is based on the number of posts that the documents appears in. Finally, we retrieve all posts related to the sampled documents to create our training posts sample. This approach ensures that our training sample contains documents which are tagged frequently as well as documents which are tagged infrequently, and reduces the bias towards documents which are only tagged once that would exist if sampling the documents uniformly at random. The resulting sample has the same distribution of documents over number of posts as the full dataset. We employ this approach of first sampling documents and then retrieving the related posts since the number of documents and the resulting size of the content data is the limiting factor which impacts recommendation speed the most in our content-based approaches. Moreover, documents don't suffer from other issues that exist when sampling users or tags and then retrieving all related posts. With users, the number of posts per user varies much more than the number of posts per document, partly due to some users using bulk imports and automatic post submission plug-ins which make them much more frequent users of the system than others. With tags, there is also much more variance in the number of posts per tag than with documents, where the issue is that tags such as ``toread", which hold no collaborative value have, a high number of related posts.

We aim to find a sample size which strikes a good balance between improving the recommendation speed of the algorithm and reducing sample bias. We want to select a sample size at which we achieve a low variation in prediction quality for different samples of the same size, and at which the increase to a larger sample is not justified by a significant increase in prediction quality. To find an appropriate sample size, we create 5 different training samples of the same size and evaluate models built on them against the test set to find the amount of variation in prediction results. We then increase the sample size to a larger value and repeat the same process, until we are confident that the sample size gives a low variation in different samples of the same size and the move to a larger sample does not significantly increase results.
We start at a sample size of 100,000 posts, and increase the number of posts by 50,000 until we are satisfied with the resulting samples.

\begin{figure}
\begin{center}
\includegraphics[scale=0.4]{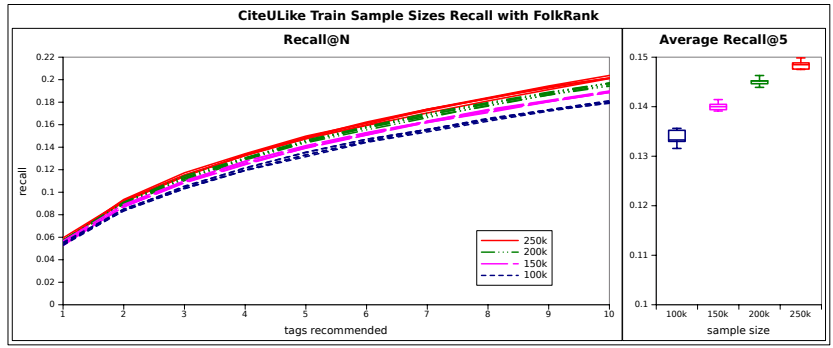}
\caption{Recall with FolkRank for Training Sample Sizes of CiteULike}
\label{fig:sample_sizes_citeulike}
\end{center}
\end{figure}

\begin{figure}
\begin{center}
\includegraphics[scale=0.4]{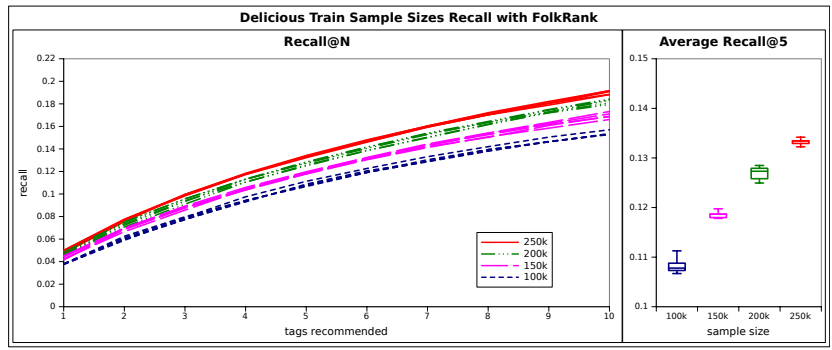}
\caption{Recall with FolkRank for Training Sample Sizes of Delicious}
\label{fig:sample_sizes_delicious}
\end{center}
\end{figure}

\begin{figure}
\begin{center}
\includegraphics[scale=0.5]{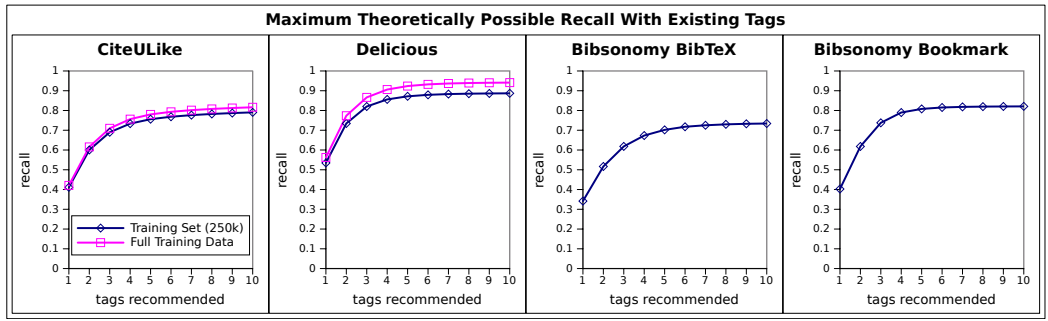}
\caption{Theoretical Maximum Recall Achievable With Existing Tags}
\label{fig:max_possible}
\end{center}
\end{figure}

Figures \ref{fig:sample_sizes_citeulike} and \ref{fig:sample_sizes_delicious} give the results with standard FolkRank for each of the examined training sample sizes on the Citelike and Delicious datasets respectively. The left side shows the recall graph of each of the individual sample runs, where runs of the same sample size are plotted in the same line style. The box plot on the right gives the average recall@5 per sample size. The more we increase the sample size, the less variation in results on samples of the same size, and the improvement in average recall is also smaller. As an outcome of this process we have found that a sample size of (roughly) 250,000 posts gives acceptable results. For BibSonomy Bookmark and BibSonomy BibTeX we do not sample the training data and use all posts from the year previous to the test time-frame, as the number of posts in these datasets is sufficiently small. The statistics of the final training and test sets used in our experiments are given in Table \ref{tab:pcore1_train_test}. The sampling does have the effect that some of the tags in the test set of CiteULike and Delicious will not be present in their respective training sets, and thus cannot be recommended successfully by any of the evaluated recommenders. In Figure \ref{fig:max_possible} we show the theoretical maximum possible recall that could be achieved on the test set of each dataset with recommending tags that exist in the training set. For CiteULike and Delicious the maximum recall is given for our training set sample and as well as the full training data. The theoretical maximum at $N$ recommended tags is calculated by assuming that for each test post $M$ correct tags are recommended at each value of $N$, where $M$ is the minimum of $N$ and the number of true tags for the test post which also exist in the training data. The extent of the problem of not including all training data tags in the samples is not too great, and we do not believe that this will impact the validity of our conclusions as all of the evaluated recommenders will suffer from this problem to the same degree. In addition to the training-test split, we create a separate evaluation split for each dataset that we use for comparison of individual methods and for parameter tuning. The evaluation test and training sets are created from the datasets prior to the two months of real test data, in the same fashion as described above.

\begin{table}
\tbl{Training and Test Set Sizes (No-Core)\label{tab:pcore1_train_test}}{
\begin{tabular}{|l|r|r|r|r|r|}
    \hline
	& \multicolumn{5}{c|}{Training Set (250k Sample)}\\ \hline
	& Posts & Tag Assignments
	& Users & Docs & Tags\\ \hline
    CiteULike & 249,968 & 1,148,011
    & 12,908 & 218,601 & 138,024\\ \hline
    Delicious & 253,890 & 566,173
    & 30,848 & 109,201 & 56,338\\ \hline
    BibS. Bookmark & 42,325 & 179,599
	& 982 & 40,679 & 24,830 \\ \hline
    BibS. BibTeX & 17,560 & 66,529
    & 1,264 & 16,360 & 17,307\\ \hline
    \hline

	& \multicolumn{5}{c|}{Test Set}\\ \hline
	& Posts & Tag Assignments
	& Users & Docs & Tags\\ \hline
    CiteULike & 76,491 & 301,779
    & 4,930 & 69,161 & 52,576\\ \hline
    Delicious & 1,746,483 & 4,431,116
    & 43,997 & 877,593 & 175,146\\ \hline
    BibS. Bookmark & 9,506 & 30,053
    & 243 & 9,425 & 4,811\\ \hline
    BibS. BibTeX & 2,843 & 10,657
    & 333 & 2,746 & 3,943\\ \hline
  \end{tabular}
  }
\end{table}

\subsection{Training and Test Set for Post-Cores Level 2}
For completeness we also evaluate our approaches on each of the datasets at post-core level 2. Post-cores at level $n$ have the constraint that each of the users, document and tags has to appear in at least $n$ posts. For each dataset, we create the post-core by iteratively removing posts where the user, document or one of the tags does not satisfy the condition that they appear in at least two posts. We then use a leave-one-out per user split to create the training and test sets by selecting the most recent post for each user and placing it in the test set. For parameter tuning we create an additional evaluation split from the resulting training data.

\section{Evaluation and Results}
\label{sec:evaluation_and_results}
In our evaluation we aim to find the best combination of our proposed approaches by answering the research questions given below. In order to achieve this we run experiments on the evaluation set where we set default values for the parameters of dampening factor ($d=0.5$) and balance in query preference weight ($b=0.5$). Having identified the best strategies, we evaluate the remaining parameters, and finally give results on the real test set with tuned parameters in Section \ref{sec:results}.

\begin{itemize}
\item[] Content Inclusion
\begin{itemize}
\item How should content be included: directly into the graph at the word level or indirectly at the document level?
\item What is the most predictive source of content: document title or fulltext content?
\item How much content should be included?
\end{itemize}
\end{itemize}

\begin{itemize}
\item[] Folksonomy Graph Model
\begin{itemize}
\item Which of the examined graph models provides the most accurate representation of the tagging data?
\end{itemize}
\end{itemize}

\begin{itemize}
\item[] Deep Folksonomy Graph
\begin{itemize}
\item Is iterative weight spreading worth the computational expense?
\item Does exploring the deeper folksonomy graph provide an improvement to tag predictions?
\end{itemize}
\end{itemize}

\subsection{Content Inclusion}

\subsubsection{Direct vs. Indirect Content Inclusion}
\begin{figure}
\begin{center}
\includegraphics[scale=0.5]{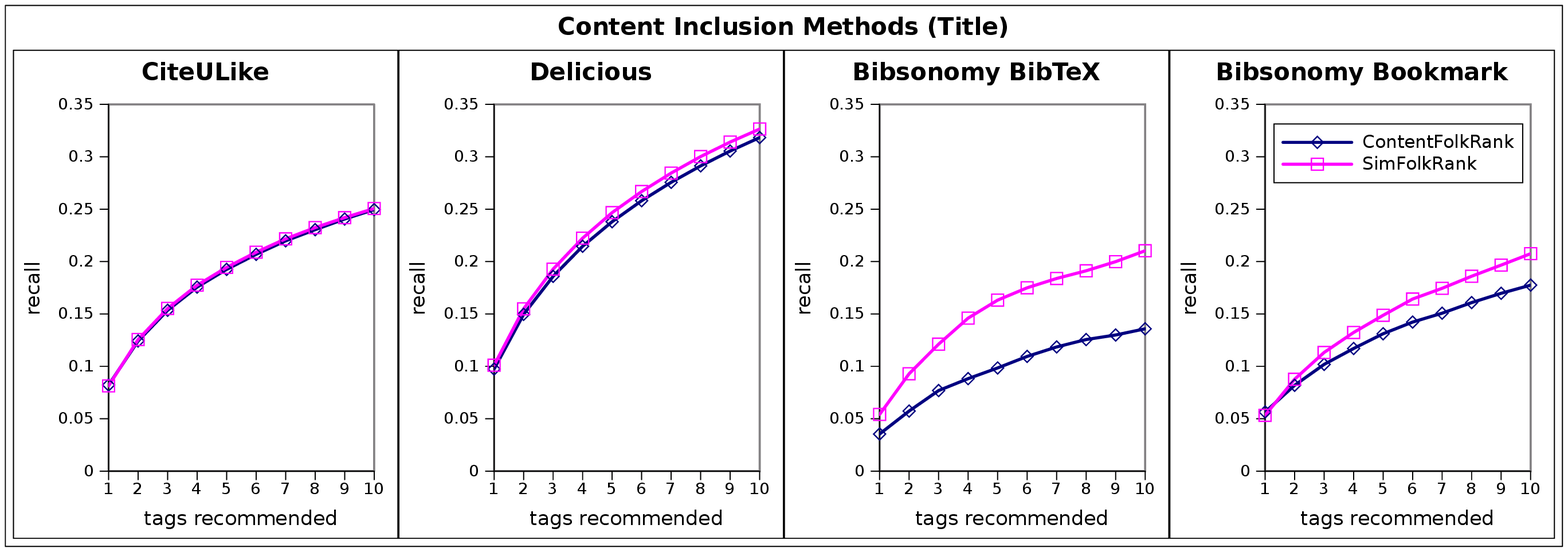}
\caption{FolkRank: Direct Content Inclusion via Word Nodes vs. Indirect Content Inclusion via Similar Documents}
\label{fig:content_inclusion}
\end{center}
\end{figure}

In Figure \ref{fig:content_inclusion} we show the results of evaluating our two methods for including content into FolkRank. On all datasets, the indirect content inclusion method of adding similar documents to the preference vector (SimFolkRank) gives better results than incorporating the document content directly into the graph (ContentFolkRank). The biggest difference is on the BibSonomy datasets, while for CiteULike the results are almost identical, with SimFolkRank performing slightly better. We assume that ContentFolkRank gives worse results due to the word nodes in the graph being connected to many more tags compared to the document nodes in the standard folksonomy graph used by SimFolkRank. The same individual word can appear in a variety of documents from different domains and thus be connected to many tags which are themselves unrelated. To accurately capture the query document several words are required in combination. The predictions generated by ContentFolkRank can be influenced by the edge configuration of individual words, which might be most connected to tags from a different domain than the query document whilst being connected to appropriate tags with less edge weight. In SimFolkRank, the similarities to training documents are calculated based on the whole representation of the query document, and in the graph each of the similar documents is likely to be connected to tags from one or a few domains only. In the larger datasets of CiteULike and Delicious the difference between the two approaches is smaller. ContentFolkRank comes close in results to SimFolkRank, but does not outperform it. This suggest that with more data the weighting methods used in ContentFolkRank, which are based on Tf-Idf scores and include a co-occurrence element, can more accurately model the query document as well as the edge weights of words in the graph. With sufficient data the outcome of ContentFolkRank is thus very similar to SimFolkRank. However, in addition to producing better results SimFolkRank is also computationally less expensive than ContentFolkRank since the ContentFolkRank graph is much larger due to the many word nodes.

\subsubsection{Content Sources}
\begin{figure}
\begin{center}
\includegraphics[scale=0.5]{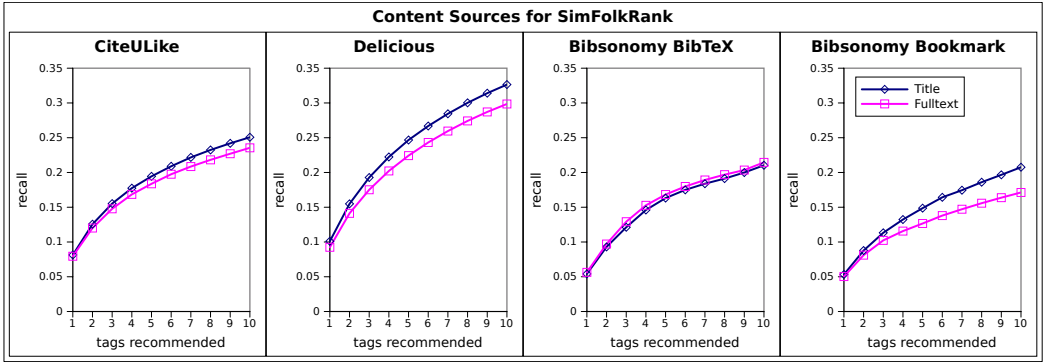}
\caption{Content Sources for SimFolkRank: Title vs. Abstract/Fulltext}
\label{fig:content_sources}
\end{center}
\end{figure}

When comparing the title and fulltext content of documents as potential document representations (Figure \ref{fig:content_sources}), the title performs better in most cases. The biggest difference is on the Delicious and BibSonomy Bookmark datasets, as here the fulltext content is the crawled page content of the bookmarked websites. In CiteULike and BibSonomy BibTeX, the fulltext representation is given by the abstract of the bookmarked research papers which we expect to be a more accurate document description. On BibSonomy BibTeX, the fulltext content actually performs slightly better than the title.

\begin{figure}
\begin{center}
\includegraphics[scale=0.5]{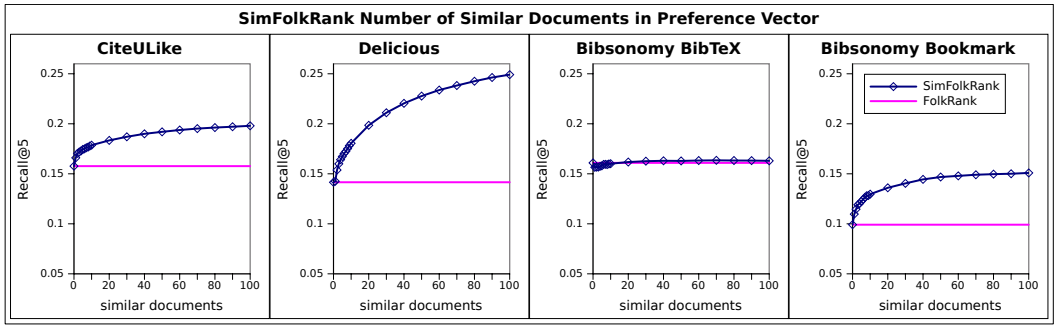}
\caption{Content Amount: Number of Similar Documents in Preference Vector of SimFolkRank}
\label{fig:SimFolkRank_numSimDocs}
\end{center}
\end{figure}

\subsubsection{Content Amount}
To evaluate the amount of content to be included we vary the number of similar documents in the preference vector of SimFolkRank and give the results in Figure \ref{fig:SimFolkRank_numSimDocs}. The content source in these experiments is the document title. The x-axis gives the number of most similar documents included in the preference vector and the y-axis is recall when recommending five tags. The left-most point and the horizontal line in each graph gives the recall@5 without including content. The results indicate that prediction results improve the more content is added, where the biggest gain is achieved by the most similar documents. The only exception is BibSonomy BibTeX where including content does not give a significant gain. Except for BibSonomy BibTeX, the shape of the plots and the fact that the results do not decrease at higher numbers of similar documents also confirms that normalised cosine similarity is an appropriate metric for measuring document similarity in our scenario.

\subsection{Graph Models}
\label{sec:eval_graph_construction}

\begin{figure}
\begin{center}
\includegraphics[scale=0.5]{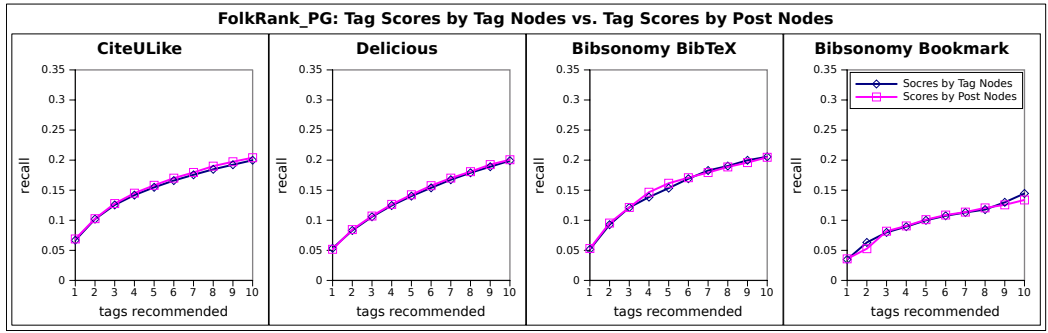}
\caption{Post Graph Scores Retrieval Method}
\label{fig:postrank_scores}
\end{center}
\end{figure}

\begin{figure}
\begin{center}
\includegraphics[scale=0.5]{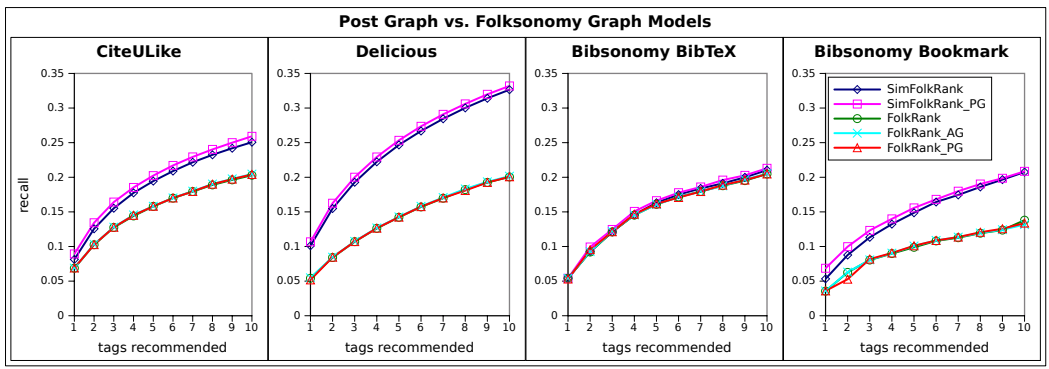}
\caption{Post Graph vs. Folksonomy Graph Models}
\label{fig:graph_construction}
\end{center}
\end{figure}

\subsubsection{Post Graph Scores Retrieval Method}
\label{sec:postrank_scores}
Before comparing the graph construction methods we first evaluate the two alternative scores retrieval methods of the Post Graph model described in Section \ref{sec:postrank}. The approach of retrieving post node weights from the graph and then calculating the tag scores based on these gives slightly better results than retrieving the tag node weights directly from the graph, although it does not seem to make a significant difference. Since it also makes sense that the number of tags in each post should not influence the scores of the tags they contain, we use the strategy of calculating tag scores from post nodes for all approaches using the Post Graph model in the subsequent experiments.

\subsubsection{Post Graph vs. Folksonomy Graph Models}
As shown in Figure \ref{fig:graph_construction}, without content data there is no real difference in results with the different models, and the folksonomy graph (FolkRank), Adapted Graph (FolkRank\_AG) and Post Graph (FolkRank\_PG) give almost identical results. However, when including content data, the Post Graph model performs consistently better than the folksonomy graph, indicated by SimFolkRank\_PG performing better than SimFolkRank across all datasets. We believe the improved results to be due to the more accurate data representation of the Post Graph model, as discussed in Section \ref{sec:graph_construction}. With more nodes in the preference vector, the implicit assumptions of the folksonomy model have a relatively greater impact on tag predictions scores and the Post Graph proves to be the more robust model.

\subsection{Weight Spreading Methods}

\subsubsection{Iterative vs. PathRank Weight Spreading}
\label{sec:eval_iterative_vs_pathrank}
\begin{figure}
\begin{center}
\includegraphics[scale=0.5]{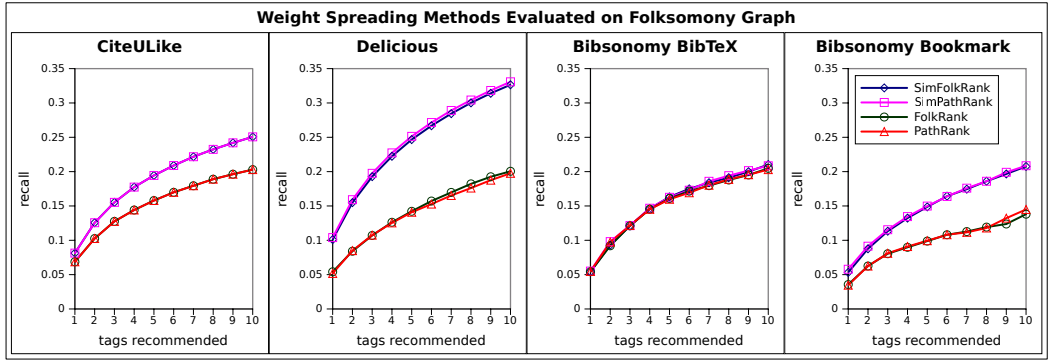}
\caption{Iterative vs. PathRank Weight Spreading on Folksonomy Graph}
\label{fig:weight_spreading_folkrank}
\end{center}
\end{figure}

\begin{figure}
\begin{center}
\includegraphics[scale=0.5]{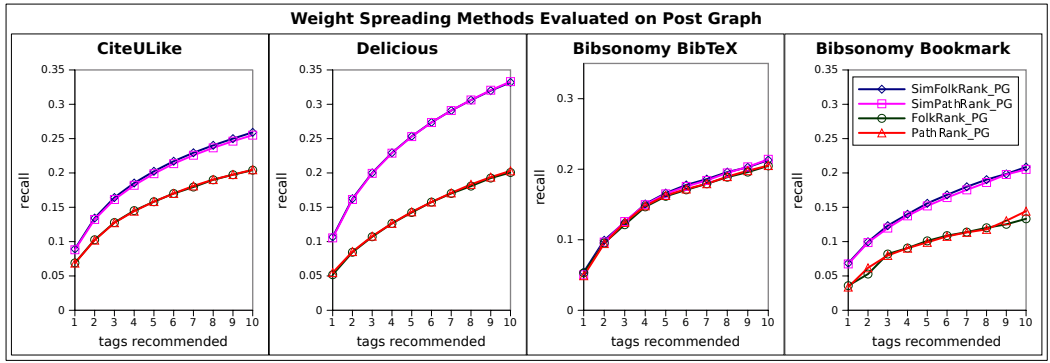}
\caption{Iterative vs. PathRank Weight Spreading on Post Graph}
\label{fig:weight_spreading_postrank}
\end{center}
\end{figure}

We compare the iterative spreading algorithm of FolkRank to our PathRank weight spreading approach on the folksonomy graph (Figure \ref{fig:weight_spreading_folkrank}) and the Post Graph model (Figure \ref{fig:weight_spreading_postrank}). The two weight spreading methods produce very similar results on both models across all datasets. However, PathRank is a much quicker weight spreading algorithm. It does not adjust the weight of each node in several iterations to find the optimal distribution of weights reflecting the overall edge connections in the graph. In other words, it does not consider the general (non-personal) importance weight of nodes which is implied by the graph structure itself. This suggests that the impact of the general importance (or authority) of nodes in the graph does not provide a significant benefit to the tag predictions, and the expensive iterative spreading of the non-personalised weights can be omitted to speed up the recommendation process. Our evaluation of the dampening factor in Section \ref{sec:tuning_d} further confirms this conclusion as the best results with FolkRank's iterative weight spreading are achieved at the lowest setting for $d$, which translates to giving the least relevance to general importance weights.

\subsubsection{Value of General Importance Weights}
\label{sec:tuning_d}
\begin{figure}
\begin{center}
\includegraphics[scale=0.5]{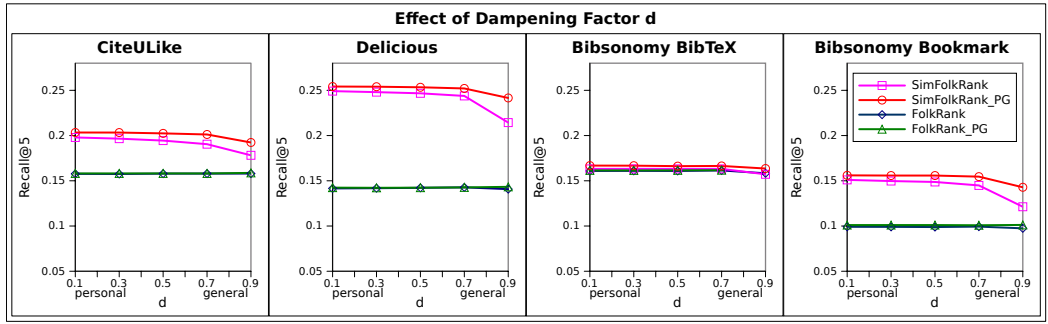}
\caption{Effect of Dampening Factor $d$ on Recall@5}
\label{fig:parameter_tuning_d}
\end{center}
\end{figure}

To examine the value of including the general node weights in the recommendation process, we evaluate different settings for the dampening factor $d$ and give our results in Figure \ref{fig:parameter_tuning_d}. Without the inclusion of content data there is not much impact on the results for the examined values of $d$. This is because without content the whole preference weight is given to a maximum of two preference nodes, the query user and document, which means that there is a huge difference in weight between the preference nodes and any one of the other nodes in the graph. Non-preference nodes, and thus general importance weights, don't have a chance to impact the predictions except for extreme values of $d$ such as 0.9, at which setting we observe a very slight decrease in results. With content data the preference weight is distributed among a maximum of 101 preference nodes, which include the query user and potentially 100 training documents similar to the query document. Here the impact of the general non-personalised weights can be observed at lower values of $d$. In all cases, the best results are achieved with setting $d$ to the lowest examined value of 0.1. This indicates that the general weights in the graph do not provide a benefit to the accuracy of tag predictions, and in fact have a negative impact when given too much relevance. We conclude that to maximise the tag prediction accuracy, $d$ should be set to the lowest value, in effect ignoring the general/non-personalised weights of nodes in the graph. With the lowest examined setting of $d=0.1$, the general weights can still act as tie-breakers for tags in the candidate set which have otherwise equal personalised weights. However, our results in the comparison with PathRank weight spreading, which does not utilise general weights, suggest that there is no significant improvement over randomly ranking tags which have equal weights. This comparison is made in the previous Section \ref{sec:eval_iterative_vs_pathrank} and in the evaluation on the real test set with tuned parameters in Section \ref{sec:results}.

\subsubsection{Predictive Value of Deep Graph}
\begin{figure}
\begin{center}
\includegraphics[scale=0.5]{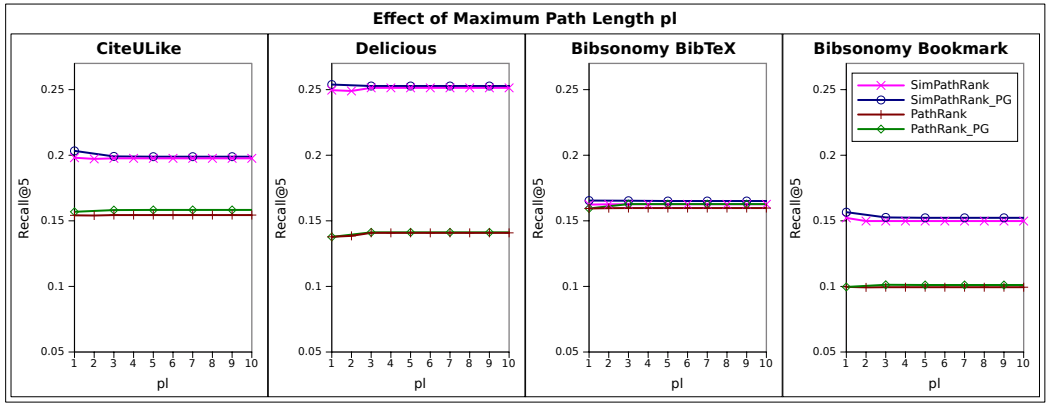}
\caption{PathRank: Effect of Different Settings of Maximum Path Length $pl$}
\label{fig:parameter_tuning_pl}
\end{center}
\end{figure}

The parameter of maximum path length $pl$ in our PathRank weight spreading approach is especially interesting since it allows us to examine the value of exploring the graph beyond the immediate neighbourhood of the query user and nodes related to the query document. We show the outcome of setting different values of $pl$ in Figure \ref{fig:parameter_tuning_pl} on the folksonomy graph and Post Graph models. The x-axis gives the value of $pl$ and the y-axis is recall@5. With the lowest setting of $pl$ only the immediate neighbourhood is explored, where as we move to the right of the x-axis longer paths are also traversed by the weight spreading algorithm. With the Post Graph model we retrieve tag scores as the sum of weights of post nodes they are connected to. Here, the next posts and thus additional tags beyond the immediate neighbourhood (of path length 1) are encountered at a path length of 3. Overall, our results suggest that there is actually not much value in considering the graph beyond the immediate neighbourhood of the preference nodes. There is a small difference that can be observed between path lengths 1 and 3 which we explore in detail below. In general, the conclusion that no significant increase is achieved is in line with our previously published results \cite{Landia2012}, where our less expensive co-occurrence recommender (exploring only the immediate neighbourhood of the query user and document) performed equally well to FolkRank. Moreover, with the PathRank weight spreading algorithm, we have now removed the other influences on the weight spreading calculation which could have cascaded or reduced the impact of the deeper graph. Even without swash-back, triangle-spreading of weights and general importance scores, the weights spread through long paths in the deep graph do not provide a significant improvement. The results indicate that the deeper graph does not provide a beneficial re-ranking of existing candidate tags in the immediate user or document neighbourhood. With the setting of $pl=1$ where only the immediate neighbourhood is considered, tag nodes which have equal weight will be ranked randomly in the final predictions. Utilising the deep graph to re-rank these tags does not significantly improve results over this random ranking.

\label{sec:tuning_pl}

Our first detailed observation is that after the first influence of the deeper graph at path length 3, we cannot observe any significant impact, positive or negative, caused by exploring longer paths. Along the lines of \cite{Wetzker2010}, this suggest that users of (broad) folksonomies have a highly personal tagging behaviour. It is thus very difficult to traverse more than a few edges in the graph and still weigh the encountered nodes in a manner relevant to the the preference node at which the path started. The only small change that can be observed is up to a path length of three. As a side note, the Post Graph model gives better results than the folksonomy graph at a path length of one. At this setting the only difference in the tag scores calculation between the two models is that for the Post Graph the tag scores are given as the sum of weights of post nodes they are connected to. As discussed in section \ref{sec:graph_construction}, this follows from the Post Graph model's assumption that the number of tags of each post should not influence tag scores, whereas the plain folksonomy graph assumes that if there are many tags in a post then each of them is less important. This again suggests that the assumptions made by the Post Graph model provide a more accurate representation of the underlying social bookmarking data.

Another interesting observation in Figure \ref{fig:parameter_tuning_pl} can be made from the results with the folksonomy graph model on the Delicious dataset. In this case there is a small improvement at a path length of 3. What is interesting here is that the increase does not occur at $pl=2$ but at $pl=3$. In the folksonomy graph, the tags found at a path length of 2 have paths of the form $u_p \rightarrow d \rightarrow t$ or $d_p \rightarrow u \rightarrow t$ from the user preference node $u_p$ or the document's preference node(s) $d_p$ respectively. Including these additional tags is conceptually similar to tag expansion via the document or user nodes related to the preference node. At a path length of 3, paths of the form $u_p \rightarrow t \rightarrow \lbrace u \vee d \rbrace  \rightarrow t$ and $d_p \rightarrow t \rightarrow \lbrace u \vee d \rbrace  \rightarrow t$ are also included which is conceptually similar to performing tag expansion by using tag-tag co-occurence measure. The small improvement in prediction accuracy seems to be due to using tag-tag co-occurence, rather than giving weight to tags which are related to non-tag nodes from the preference node's immediate neighbourhood. On the BibSonomy Bookmark dataset we can observe a small decrease at $pl=2$ when including content with SimPathRank. With the Post Graph model and content (SimPathRank\_PG), there is also a decrease on BibSonomy as well as CiteULike when going from $pl=1$ to $pl=3$. As there are no paths with length 2 leading to additional tags in the Post Graph, the influence of tag expansion both via non-tag nodes and tag-tag co-occurrence is included at the same time at $pl=3$. It seems to be the case that tag expansion via non-tag nodes decreases results. Along the lines of our discussion in Section \ref{sec:weight_spreading_discussion}, this seems to suggest that tags found related to non-tag nodes of the preference node but not directly connected to the preference node itself should not be given an increased weight. As they seem to worsen results it might be appropriate to decrease their weight instead. This suggest a potential that negative feedback could be extracted via a more complex analysis of the graph, which we intend to investigate in the future.

Overall, we conclude that spreading weight into the deeper graph does not provide a significant benefit to tag recommendations and can in some cases even harm prediction scores. The only increase in scores is given by spreading weight from tags to further tag nodes, essentially performing a tag set expansion via tag-tag co-occurrence. Given the complete graph model this is very difficult to separate from expanding the tag set via non-tag nodes, which seems to decrease prediction accuracy. To still utilise the tag-tag co-occurrence data we believe that separate approaches which directly model the tag-tag relationships would be more appropriate and produce better results. However, even though the assumptions made by conventional positive-reinforcement weight spreading methods do not seem to hold for the social bookmarking domain, some useful information could potentially be gained from the deep folksonomy graph by different approaches. A rule-driven analysis of small subsections of the graph could be used to make deductions about implied negative feedback, to either aid the recommendation process directly or to improve the accuracy of a tag-tag similarity metric by including negative scores.

\subsection{Tuning of Remaining Parameters}
\subsubsection{Balance in Preference Weight Between Query User and Query Document}
\label{sec:tuning_b}

\begin{figure}
\begin{center}
\includegraphics[scale=0.5]{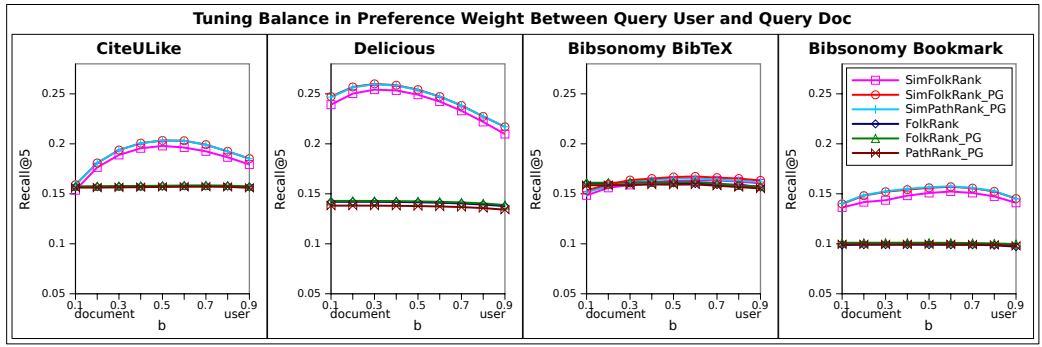}
\caption{Parameter Tuning of Balance $b$ Between Query User and Query Document}
\label{fig:parameter_tuning_b}
\end{center}
\end{figure}

In Figure \ref{fig:parameter_tuning_b} we present the results for different settings of $b$, which determines the balance in preference weight between the query user and the query document. Once again there is not much difference in results without including content data (FolkRank, FolkRank\_PG, PathRank\_PG). Since most of the query documents in the test sets are new, the preference vector without content will only include the query user in the majority of cases. For the cases where the document does exist in the graph, and thus will be included in the preference vector, each of the tags connected to the query document will usually receive more weight than each of the tags connected to the query user since users are usually connected to many more tags than documents are. The tags connected to the query user only have a chance to outweigh the tags connected to the query document for high values of $b$, at which settings we see a slight decrease in results. However, with content data (SimFolkRank, SimFolkRank\_PG, SimPathRank\_PG) the preference vector contains the query user as well as several documents related to the query document and we can clearly observe the impact of $b$. The results confirm that there is value in introducing the parameter $b$ to explicitly set this balance instead of using the strategy of the original FolkRank algorithm of setting $b=\frac{|U|}{|U|+|D|}$, which results in values lower than 0.1 for all of the datasets except Delicious where it would be 0.2. The best results are achieved with setting $b$ to 0.5 for CiteULike, 0.3 for Delicious, and 0.6 for both BibSonomy Bookmark and Bibsonomy BibTeX.

\subsection{Results on Test Set}
\label{sec:results}
Here we present our final results with our best approaches and with tuned parameters on the test set of each of the datasets. The content source in all of the content-aware approaches is the document title. For approaches using FolkRank's iterative weight spreading the dampening factor is set to $d=0.1$, and for approaches using PathRank the maximum path length is set to $pl=1$. The balance $b$ in preference weight is set per dataset to the best value that was found in the parameter tuning runs.

\begin{figure}
\begin{center}
\includegraphics[scale=0.45]{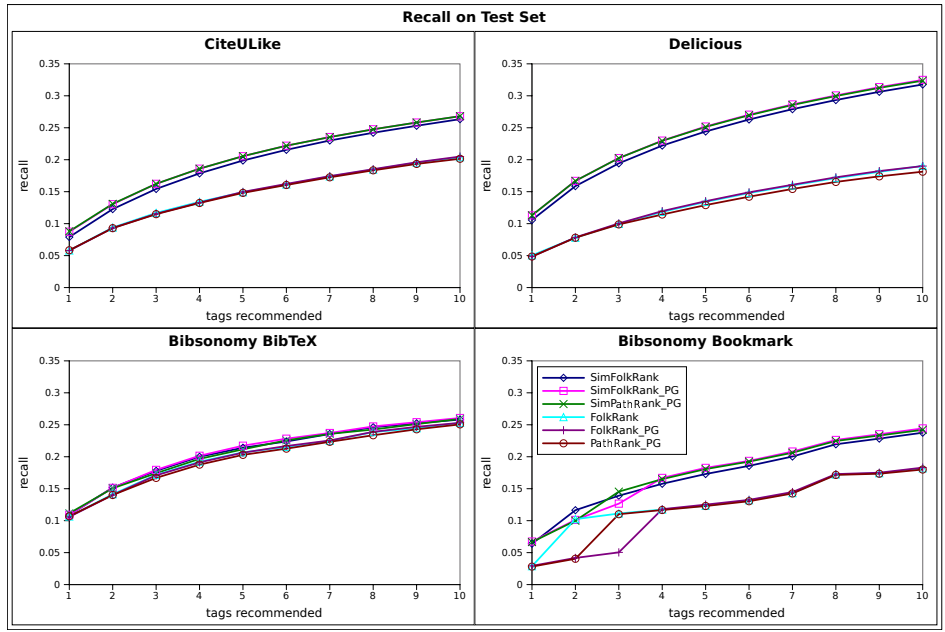}
\caption{Recall on Test Set with Tuned Parameters}
\label{fig:results_test_set}
\end{center}
\end{figure}

Figures \ref{fig:results_test_set} and \ref{fig:results_test_set} show the recall and F1 respectively, on the test set for each of the datasets with tuned parameters. Including content into the recommendation process provides a significant increase in results. The results on the test set are in line with our previous conclusions on the evaluation set. SimFolkRank\_PG produces better results than SimFolkRank over all datasets, suggesting that the Post Graph is a more accurate model of the tagging data than the folksonomy graph. Furthermore, PathRank\_PG and SimPathRank\_PG give almost equivalent results to FolkRank\_PG and SimFolkRank\_PG respectively which suggests that the iterative computation and general importance weights in FolkRank's weight spreading approach do not provide a significant benefit to tag predictions. While producing comparable results, the PathRank weight spreading method is much less computationally expensive. Furthermore, the results with SimPathRank\_PG, which is among the best recommenders across all datasets, are achieved with a parameter setting of $pl=1$. At this setting only the immediate neighbourhood of preference nodes is considered. None of the approaches improve results by utilising the deep graph over SimPathRank\_PG with $pl=1$ which is essentially a user-tag and document-tag co-occurrence recommender at this setting.
The results on BibSonomy Bookmark without including content data are due to the fact that a large portion of the test posts in BibSonomy Bookmark contain new users as well as new documents. For these test posts the algorithms which don't include content data (FolkRank, FolkRank\_PG and PathRank\_PG) default to recommending the overall highly-ranked tags in the graph without personalisation. The three approaches have different rankings for the top three tags in the general recommendations which leads to the results shown.

\begin{figure}
\begin{center}
\includegraphics[scale=0.45]{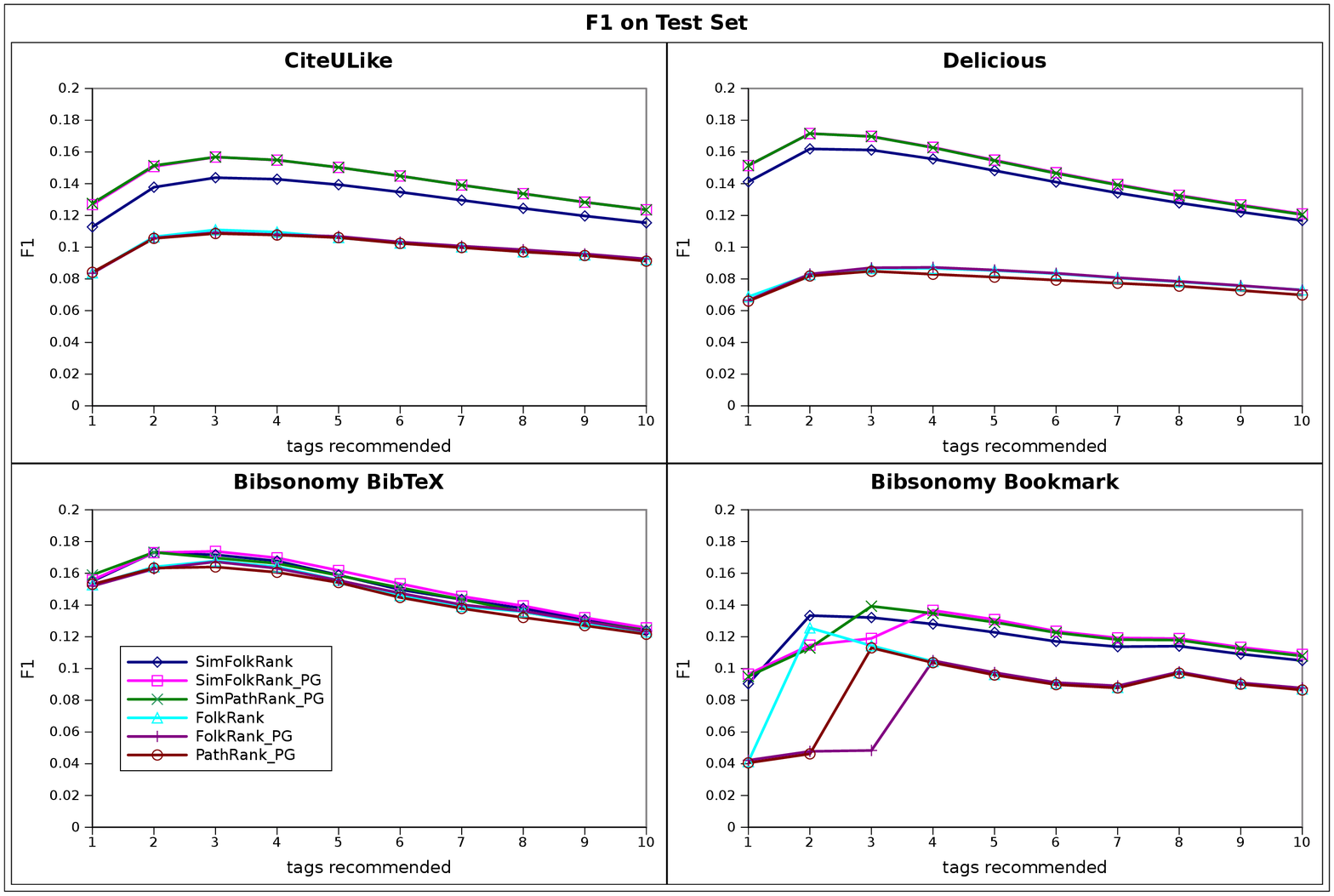}
\caption{F1 on Test Set with Tuned Parameters}
\label{fig:results_test_set_f1}
\end{center}
\end{figure}

\subsection{Post-Core 2}

\begin{figure}
\begin{center}
\includegraphics[scale=0.5]{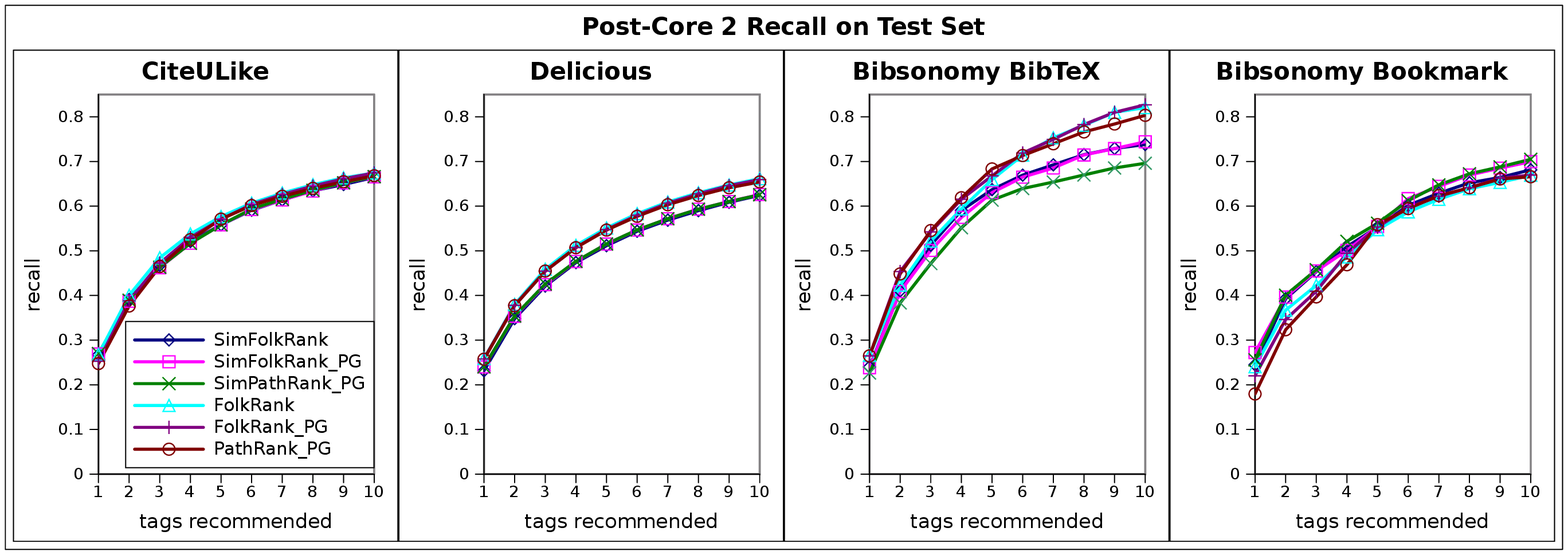}
\caption{Post-Core 2 Recall on Test Set with Tuned Parameters}
\label{fig:results_test_set_pc2}
\end{center}
\end{figure}

\begin{figure}
\begin{center}
\includegraphics[scale=0.5]{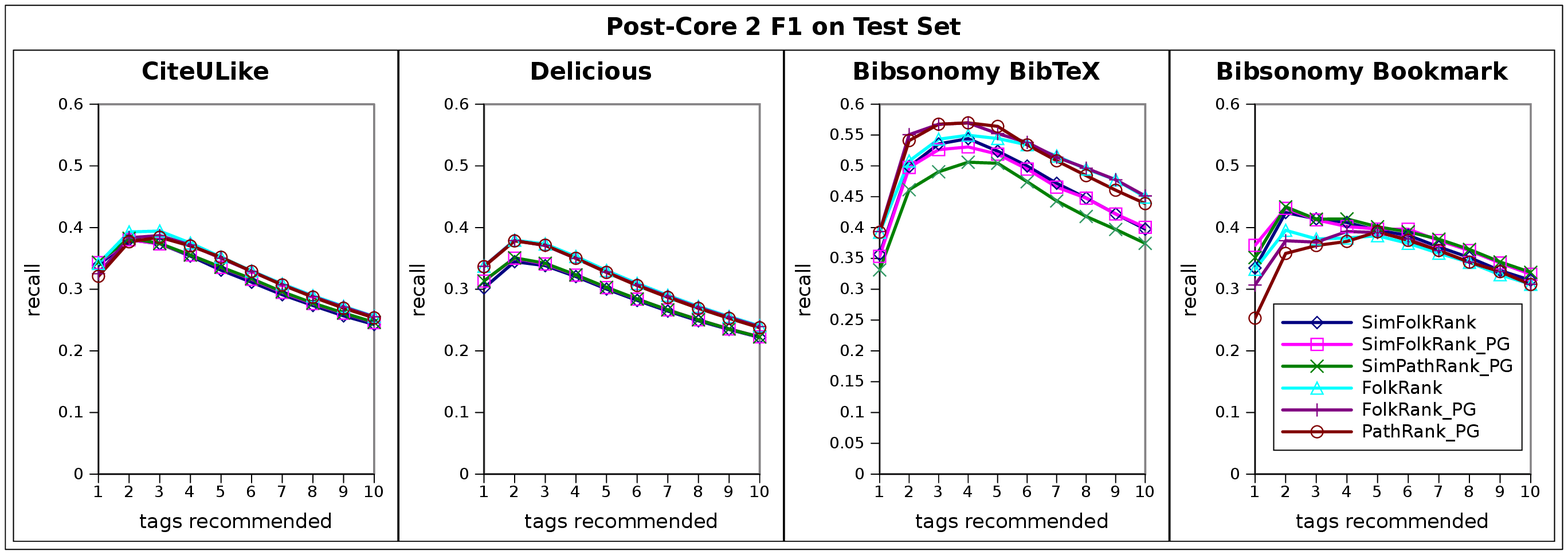}
\caption{Post-Core 2 F1 on Test Set with Tuned Parameters}
\label{fig:results_test_set_pc2_f1}
\end{center}
\end{figure}

For completeness, we show the recall and F1 for each of the datasets' post-cores at level 2 in Figures \ref{fig:results_test_set_pc2} and \ref{fig:results_test_set_pc2_f1} respectively. The results with all methods are very similar except on the smaller Bibsonomy datasets. However, there is no improvement with including content data. In fact, including content data makes results worse in most cases. Since (almost) all of the documents in the test set for post-cores also exist in the training data, they all have previously assigned tags available which can be recommended. There is no need to additionally include similar documents in the preference vector as well since the exact query document exists in the graph. Adding the content in this case has a negative effect on tag predictions. This is an interesting result and suggests that the best strategy for the future might be to only include the content if the query document does not exist in the training data, for experiments on post-core 2 as well as unpruned datasets.

\section{Conclusion and Future Work}
\label{sec:conclusion}
In this paper we have presented novel adaptations and extensions to FolkRank and conducted an in-depth analysis of the accuracy of the folksonomy graph model, the iterative weight spreading algorithm of FolkRank and the value of exploring the deep folksonomy graph. The extension of FolkRank with content data resulted in a significant increase in tag recommendation accuracy and addressed the new item problem in tag recommendation, as well as providing further insight into the FolkRank algorithm. As part of our examination of the folksonomy graph structure, we have proposed an improved model which captures the tagging data more accurately and produces better tag recommendation results. In our analysis of the iterative weight spreading method of FolkRank, we have shown that the general un-personalised node weights do not provide a positive impact on tag recommendations, and if given too much relevance hurt the accuracy of the algorithm. Since the general node weights are one of the main reasons for FolkRank's high complexity, we think it is an important finding that they can be safely omitted. Furthermore, we have shown that a simpler weight spreading algorithm, PathRank, which works in a similar manner to breadth-first search, produces comparable results to the much more complex iterative weight spreading algorithm employed by FolkRank while being computationally less expensive. The most intriguing result of our analysis was that even though both FolkRank's iterative weight spreading and our simpler PathRank spreading algorithm have the potential to utilise the deep folksonomy graph, they do not benefit from doing so in practice. Moreover, we have presented an in-depth discussion as well as a direct evaluation of the value of exploring the deep folksonomy graph. We conclude that exploring the graph beyond the immediate neighbourhood of the query nodes with conventional weight spreading methods does not provide a significant increase in tag recommendation accuracy and can in some cases even hurt recommendations. The assumption that closeness in the graph always implies a positive relationship does not hold beyond the immediate neighbourhood of nodes in social tagging graphs. This suggests that the foundation of graph-based recommenders (and to a lesser extent collaborative filtering), which are traditionally applied to two-dimensional datasets, does not apply to the three-dimensional user-document-tag relationships found in social tagging data. In summary our main conclusions are as follows.

\begin{itemize}
\item[] Content Inclusion in Tag Recommendation
\begin{itemize}
\item Including content into the recommendation process addresses the new document problem and significantly increases results on full/unpruned datasets.
\item The title of documents is a better content source and provides a more accurate description of documents than the fulltext content.
\item Including content at the document level produces a more accurate recommender than including content at the word level and constructing user-word and word-tag relationships, especially for smaller sized social tagging datasets.
\end{itemize}
\end{itemize}

\begin{itemize}
\item[] Folksonomy Graph Model
\begin{itemize}
\item Explicitly including post-membership information into the graph provides a model which makes more accurate assumptions about the relationships in the tagging data and produces improved results over the traditional folksonomy model.
\end{itemize}
\end{itemize}

\begin{itemize}
\item[] Deep Graph Exploration
\begin{itemize}
\item General importance/authority scores, which make iterative weight spreading computationally expensive, do not provide an improvement to the accuracy of tag recommendations and can be omitted to reduce complexity.
\item The expensive exploration of the deep tagging data graph with conventional weight spreading methods does not provide an improvement to tag recommendations and can in some cases decrease results.
\item The assumption that closeness in the graph always implies a positive relationship does not hold in social tagging datasets beyond the immediate neighbourhood of nodes.
\end{itemize}
\end{itemize}

In the future we plan to further explore methods to leverage the potential benefit of including the information contained in the deep folksonomy graph for tag recommendation. We think that by using rule-based methods which analyse smaller subgraphs of the folksonomy, implicit negative feedback could be extracted. This could be used to include negative scores in user-tag and especially document-tag relationships in order to reduce the scores of tags which are likely to be incorrect for a specific user or document. Moreover, the negative feedback could be incorporated into tag-tag similarity measures to make these more accurate. Another interesting research direction are the sampling methods used in tag recommendation. As social bookmarking websites and tagging datasets get larger, it is becoming infeasible to build models on and analyse all of the training data, especially with methods which examine complex relationships in the data. We plan to further explore this problem and evaluate different sampling methods in their ability to produce unbiased and predictive samples of training data.

\bibliographystyle{ACM-Reference-Format-Journals}

\begin{thebibliography}{00}

\ifx \showCODEN    \undefined \def \showCODEN     #1{\unskip}     \fi
\ifx \showDOI      \undefined \def \showDOI       #1{{\tt DOI:}\penalty0{#1}\ }
  \fi
\ifx \showISBNx    \undefined \def \showISBNx     #1{\unskip}     \fi
\ifx \showISBNxiii \undefined \def \showISBNxiii  #1{\unskip}     \fi
\ifx \showISSN     \undefined \def \showISSN      #1{\unskip}     \fi
\ifx \showLCCN     \undefined \def \showLCCN      #1{\unskip}     \fi
\ifx \shownote     \undefined \def \shownote      #1{#1}          \fi
\ifx \showarticletitle \undefined \def \showarticletitle #1{#1}   \fi
\ifx \showURL      \undefined \def \showURL       #1{#1}          \fi

\bibitem[\protect\citeauthoryear{Brin and Page}{Brin and Page}{1998}]%
        {Brin98}
{Sergey Brin} {and} {Lawrence Page}. 1998.
\newblock \showarticletitle{The anatomy of a large-scale hypertextual Web
  search engine}, In Proceedings of the 7th International Conference on World
  Wide Web 7. {\em Computer Networks and ISDN Systems\/} {30}, 1-7 (April
  1998), 107--117.
\newblock
\showISSN{01697552}
\showDOI{%
\url{http://dx.doi.org/10.1016/S0169-7552(98)00110-X}}


\bibitem[\protect\citeauthoryear{Eisterlehner, Hotho, and
  J{\"a}schke}{Eisterlehner et~al\mbox{.}}{2009}]%
        {proceedings_dc09}
{Folke Eisterlehner}, {Andreas Hotho}, {and} {Robert J{\"a}schke} (Eds.). 2009.
\newblock {\em Proceedings of ECML PKDD Discovery Challenge 2009, Bled,
  Slovenia, September, 2009}. CEUR-WS.org, Vol. 497.
\newblock


\bibitem[\protect\citeauthoryear{Gemmell, Schimoler, Ramezani, and
  Mobasher}{Gemmell et~al\mbox{.}}{2009}]%
        {Gemmell2009}
{Jonathan Gemmell}, {Thomas Schimoler}, {Maryam Ramezani}, {and} {Bamshad
  Mobasher}. 2009.
\newblock \showarticletitle{Adapting K-Nearest Neighbor for Tag Recommendation
  in Folksonomies}. In {\em Proceedings of the 7th Workshop on Intelligent
  Techniques for Web Personalization and Recommender Systems}.
\newblock


\bibitem[\protect\citeauthoryear{Heymann, Ramage, and Garcia-Molina}{Heymann
  et~al\mbox{.}}{2008}]%
        {Heymann2008}
{Paul Heymann}, {Daniel Ramage}, {and} {Hector Garcia-Molina}. 2008.
\newblock \showarticletitle{Social tag prediction}. In {\em SIGIR '08:
  Proceedings of the 31st Annual International ACM SIGIR Conference on Research
  and Development in Information Retrieval}. ACM, New York, NY, USA, 531--538.
\newblock
\showISBNx{978-1-60558-164-4}
\showDOI{%
\url{http://dx.doi.org/10.1145/1390334.1390425}}


\bibitem[\protect\citeauthoryear{Hotho, J{\"a}schke, Schmitz, and Stumme}{Hotho
  et~al\mbox{.}}{2006}]%
        {Hotho2006}
{Andreas Hotho}, {Robert J{\"a}schke}, {Christoph Schmitz}, {and} {Gerd
  Stumme}. 2006.
\newblock \showarticletitle{Information Retrieval in Folksonomies: Search and
  Ranking}. In {\em The Semantic Web: Research and Applications} {\em (Lecture
  Notes in Computer Science)}, Vol. 4011. Springer, 411--426.
\newblock
\showISBNx{978-3-540-34544-2}
\showDOI{%
\url{http://dx.doi.org/10.1007/11762256_31}}


\bibitem[\protect\citeauthoryear{Hulth}{Hulth}{2003}]%
        {Hulth2003}
{Anette Hulth}. 2003.
\newblock \showarticletitle{Improved automatic keyword extraction given more
  linguistic knowledge}. In {\em Proceedings of the 2003 Conference on
  Empirical Methods in Natural Language Processing}. Association for
  Computational Linguistics, Morristown, NJ, USA, 216--223.
\newblock
\showDOI{%
\url{http://dx.doi.org/10.3115/1119355.1119383}}


\bibitem[\protect\citeauthoryear{J\"{a}schke, Marinho, Hotho, Schmidt-Thieme,
  and Stumme}{J\"{a}schke et~al\mbox{.}}{2007}]%
        {Jaeschke2007}
{Robert J\"{a}schke}, {Leandro~Balby Marinho}, {Andreas Hotho}, {Lars
  Schmidt-Thieme}, {and} {Gerd Stumme}. 2007.
\newblock \showarticletitle{Tag Recommendations in Folksonomies}. In {\em
  Knowledge Discovery in Databases: PKDD 2007, 11th European Conference on
  Principles and Practice of Knowledge Discovery in Databases}. 506--514.
\newblock


\bibitem[\protect\citeauthoryear{Kim and El~Saddik}{Kim and El~Saddik}{2011}]%
        {Kim2011}
{Heung-Nam Kim} {and} {Abdulmotaleb El~Saddik}. 2011.
\newblock \showarticletitle{Personalized PageRank vectors for tag
  recommendations: Inside FolkRank}. In {\em Proceedings of the fifth ACM
  conference on Recommender systems} {\em (RecSys '11)}. ACM, New York, NY,
  USA, 45--52.
\newblock
\showISBNx{978-1-4503-0683-6}
\showDOI{%
\url{http://dx.doi.org/10.1145/2043932.2043945}}


\bibitem[\protect\citeauthoryear{Landia, Anand, Hotho, J\"{a}schke, Doerfel,
  and Mitzlaff}{Landia et~al\mbox{.}}{2012}]%
        {Landia2012}
{Nikolas Landia}, {Sarabjot~Singh Anand}, {Andreas Hotho}, {Robert
  J\"{a}schke}, {Stephan Doerfel}, {and} {Folke Mitzlaff}. 2012.
\newblock \showarticletitle{Extending FolkRank with content data}. In {\em
  Proceedings of the 4th ACM RecSys workshop on Recommender Systems and the
  Social Web} {\em (RSWeb '12)}. ACM, New York, NY, USA, 1--8.
\newblock
\showISBNx{978-1-4503-1638-5}
\showDOI{%
\url{http://dx.doi.org/10.1145/2365934.2365936}}


\bibitem[\protect\citeauthoryear{Lipczak, Hu, Kollet, and Milios}{Lipczak
  et~al\mbox{.}}{2009}]%
        {Lipczak2009}
{Marek Lipczak}, {Yeming Hu}, {Yael Kollet}, {and} {Evangelos Milios}. 2009.
\newblock \showarticletitle{Tag Sources for Recommendation in Collaborative
  Tagging Systems}. In {\em Proceedings of the ECML/PKDD 2009 Discovery
  Challenge Workshop}. 157--172.
\newblock


\bibitem[\protect\citeauthoryear{Lipczak and Milios}{Lipczak and
  Milios}{2010}]%
        {Lipczak2010b}
{Marek Lipczak} {and} {Evangelos Milios}. 2010.
\newblock \showarticletitle{The impact of resource title on tags in
  collaborative tagging systems}. In {\em HT '10: Proceedings of the 21st ACM
  Conference on Hypertext and Hypermedia}. ACM, New York, NY, USA, 179--188.
\newblock
\showISBNx{978-1-4503-0041-4}
\showDOI{%
\url{http://dx.doi.org/10.1145/1810617.1810648}}


\bibitem[\protect\citeauthoryear{Lipczak and Milios}{Lipczak and
  Milios}{2011}]%
        {Lipczak2011}
{Marek Lipczak} {and} {Evangelos Milios}. 2011.
\newblock \showarticletitle{Efficient Tag Recommendation for Real-Life Data}.
\newblock {\em ACM Trans. Intell. Syst. Technol.\/} {3}, 1, Article 2 (Oct.
  2011), 21 pages.
\newblock
\showISSN{2157-6904}
\showDOI{%
\url{http://dx.doi.org/10.1145/2036264.2036266}}


\bibitem[\protect\citeauthoryear{Liu, Pennell, Liu, and Liu}{Liu
  et~al\mbox{.}}{2009}]%
        {Liu2009}
{Feifan Liu}, {Deana Pennell}, {Fei Liu}, {and} {Yang Liu}. 2009.
\newblock \showarticletitle{Unsupervised approaches for automatic keyword
  extraction using meeting transcripts}. In {\em NAACL '09: Proceedings of
  Human Language Technologies: The 2009 Annual Conference of the North American
  Chapter of the Association for Computational Linguistics}. Association for
  Computational Linguistics, Morristown, NJ, USA, 620--628.
\newblock
\showISBNx{978-1-932432-41-1}


\bibitem[\protect\citeauthoryear{Matsuo and Ishizuka}{Matsuo and
  Ishizuka}{2004}]%
        {Matsuo2004}
{Y. Matsuo} {and} {M. Ishizuka}. 2004.
\newblock \showarticletitle{Keyword Extraction From A Single Document Using
  Word Co-Occurrence Statistical Information}.
\newblock {\em International Journal on Artificial Intelligence Tools\/}  {13}
  (2004), 157--170.
\newblock


\bibitem[\protect\citeauthoryear{Ramezani, Gemmell, Schimoler, and
  Mobasher}{Ramezani et~al\mbox{.}}{2010}]%
        {Ramezani2010}
{Maryam Ramezani}, {Jonathan Gemmell}, {Thomas Schimoler}, {and} {Bamshad
  Mobasher}. 2010.
\newblock \showarticletitle{Improving Link Analysis for Tag Recommendation in
  Folksonomies}. In {\em Proceedings of the 2nd Recommender Systems and the
  Social Web Workshop at RecSys '10}. 39--45.
\newblock


\bibitem[\protect\citeauthoryear{Rendle, Balby~Marinho, Nanopoulos, and
  Schmidt-Thieme}{Rendle et~al\mbox{.}}{2009}]%
        {Rendle2009}
{Steffen Rendle}, {Leandro Balby~Marinho}, {Alexandros Nanopoulos}, {and} {Lars
  Schmidt-Thieme}. 2009.
\newblock \showarticletitle{Learning optimal ranking with tensor factorization
  for tag recommendation}. In {\em Proceedings of the 15th ACM SIGKDD
  International Conference on Knowledge Discovery and Data Mining} {\em (KDD
  '09)}. New York, NY, USA, 727--736.
\newblock
\showISBNx{978-1-60558-495-9}
\showDOI{%
\url{http://dx.doi.org/10.1145/1557019.1557100}}


\bibitem[\protect\citeauthoryear{Renz, Ficzay, and Hitzler}{Renz
  et~al\mbox{.}}{2003}]%
        {Renz2003}
{Ingrid Renz}, {Andrea Ficzay}, {and} {Holger Hitzler}. 2003.
\newblock \showarticletitle{Keyword Extraction for Text Characterization}. In
  {\em NLDB 2003: Natural Language Processing and Information Systems, 8th
  International Conference on Applications of Natural Language to Information
  Systems, June 2003, Burg (Spreewald), Germany}. 228--234.
\newblock


\bibitem[\protect\citeauthoryear{Song, Zhuang, Li, Zhao, Li, Lee, and
  Giles}{Song et~al\mbox{.}}{2008}]%
        {Song2008}
{Yang Song}, {Ziming Zhuang}, {Huajing Li}, {Qiankun Zhao}, {Jia Li},
  {Wang-Chien Lee}, {and} {C.~Lee Giles}. 2008.
\newblock \showarticletitle{Real-time Automatic Tag Recommendation}. In {\em
  SIGIR '08: Proceedings of the 31st Annual International ACM SIGIR Conference
  on Research and Development in Information Retrieval}. 515--522.
\newblock


\bibitem[\protect\citeauthoryear{Symeonidis, Nanopoulos, and
  Manolopoulos}{Symeonidis et~al\mbox{.}}{2008}]%
        {Symeonidis2008}
{Panagiotis Symeonidis}, {Alexandros Nanopoulos}, {and} {Yannis Manolopoulos}.
  2008.
\newblock \showarticletitle{Tag recommendations based on tensor dimensionality
  reduction}. In {\em Proceedings of the 2008 ACM Conference on Recommender
  Systems} {\em (RecSys '08)}. ACM, New York, NY, USA, 43--50.
\newblock
\showISBNx{978-1-60558-093-7}
\showDOI{%
\url{http://dx.doi.org/10.1145/1454008.1454017}}


\bibitem[\protect\citeauthoryear{Wetzker, Zimmermann, Bauckhage, and
  Albayrak}{Wetzker et~al\mbox{.}}{2010}]%
        {Wetzker2010}
{Robert Wetzker}, {Carsten Zimmermann}, {Christian Bauckhage}, {and} {Sahin
  Albayrak}. 2010.
\newblock \showarticletitle{{I tag, you tag: translating tags for advanced user
  models}}. In {\em Proceedings of the third ACM international conference on
  Web search and data mining} {\em (WSDM '10)}. ACM, New York, NY, USA, 71--80.
\newblock
\showISBNx{978-1-60558-889-6}
\showDOI{%
\url{http://dx.doi.org/10.1145/1718487.1718497}}


\bibitem[\protect\citeauthoryear{Witten, Paynter, Frank, Gutwin, and
  Nevill-Manning}{Witten et~al\mbox{.}}{1999}]%
        {Witten1999}
{Ian~H. Witten}, {Gordon~W. Paynter}, {Eibe Frank}, {Carl Gutwin}, {and}
  {Craig~G. Nevill-Manning}. 1999.
\newblock \showarticletitle{KEA: Practical Automatic Keyphrase Extraction.}. In
  {\em Proceedings of the 4th ACM Conference on Digital Libraries, August
  11-14, 1999, Berkeley, CA, USA}. ACM, 254--255.
\newblock


\bibitem[\protect\citeauthoryear{Xu, Fu, Mao, and Su}{Xu et~al\mbox{.}}{2006}]%
        {Xu2006}
{Zhichen Xu}, {Yun Fu}, {Jianchang Mao}, {and} {Difu Su}. 2006.
\newblock \showarticletitle{Towards the Semantic Web: Collaborative Tag
  Suggestions}. In {\em Proceedings of the Collaborative Web Tagging Workshop
  at WWW 2006}. Edinburgh, Scotland.
\newblock


\bibitem[\protect\citeauthoryear{Zhang, Zincir-Heywood, and Milios}{Zhang
  et~al\mbox{.}}{2004}]%
        {Zhang2004}
{Yongzheng Zhang}, {Nur Zincir-Heywood}, {and} {Evangelos Milios}. 2004.
\newblock \showarticletitle{World wide web site summarization}.
\newblock {\em Web Intelli. and Agent Sys.\/} {2}, 1 (2004), 39--53.
\newblock
\showISSN{1570-1263}


\end{thebibliography}


\end{document}